\documentclass[twocolumn,showpacs,preprintnumbers,amsmath,amssymb]{revtex4}
\usepackage[pdftex]{graphicx}
\usepackage{dcolumn}
\usepackage{bm}
\usepackage[pdftex,colorlinks=true,pdfstartview=Fit]{hyperref}

\begin{document}

\title{How \textit{Xenopus laevis} embryos replicate reliably: investigating the random-completion problem}

\author{Scott Cheng-Hsin Yang}
\email{scotty@sfu.ca}

\author{John Bechhoefer}
\affiliation{Department of Physics, Simon Fraser University, Burnaby, B.C., V5A 1S6, Canada}

\date{\today}

\begin{abstract}
DNA synthesis in \textit{Xenopus} frog embryos initiates stochastically in time at many sites (origins) along the chromosome. Stochastic initiation implies fluctuations in the time to complete and may lead to cell death if replication takes longer than the cell cycle time ($\approx 25$ min). Surprisingly, although the typical replication time is about 20 min, \textit{in vivo} experiments show that replication fails to complete only about 1 in 300 times. How is replication timing accurately controlled despite the stochasticity? Biologists have proposed two solutions to this ``random-completion problem." The first solution uses randomly located origins but increases their rate of initiation as S phase proceeds, while the second uses regularly spaced origins. In this paper, we investigate the random-completion problem using a type of model first developed to describe the kinetics of first-order phase transitions. Using methods from the field of extreme-value statistics, we derive the distribution of replication-completion times for a finite genome. We then argue that the biologists' first solution to the problem is not only consistent with experiment but also nearly optimizes the use of replicative proteins. We also show that spatial regularity in origin placement does not alter significantly the distribution of replication times and, thus, is not needed for the control of replication timing.

\end{abstract}
\pacs{87.15.A-, 87.14.G-, 87.17.Ee, 87.15.Ya} 

\maketitle

\section{\label{sec:intro}Introduction}

DNA replication is an important yet complicated process that requires not only accurate and efficient DNA synthesis but also genome-wide coordination among replicative proteins \cite{herrick08}. In a time that can be as short as a few minutes, all of a cell's $\mathcal{O}(10^9)$ bases of DNA must be replicated once and only once \cite{diffley96, arias07}. Unfaithful and uncontrolled replication of the genome --- for example, mis-replication, partial replication, and re-replication --- can lead to chromosomal instability that activates programmed cell death or oncogenes \cite{hensey97, micco06}. Over the past few decades, significant advances have been made in identifying the molecular basis of DNA repair and re-replication prevention \cite{sancar04, arias07}. On the other hand, it is only in the last few years that large amounts of data on the genome-wide coordination have become available. In particular, a technique called molecular combing has been used to examine the replication state of large fractions of the genome by controlled stretching of fluorescently labeled replicated and unreplicated regions onto a substrate \cite{bensimon94, herrick99}.

Many of the molecular-combing experiments have been done on embryos of the South African clawed frog, \textit{Xenopus laevis} \cite{herrick00, marheineke01, herrick02}. The detailed kinetics of replication revealed a particularly interesting scenario where stochastic effects play an important role in the DNA replication process \cite{herrick00, hyrien93}. In previous work, we mapped the stochastic replication process onto a one-dimensional nucleation-and-growth process and modeled the detailed kinetics of replication seen in molecular-combing experiments \cite{herrick02, jun05a, jun05b}. In a recent letter, we extended the model to quantitatively address a generalized version of the ``random-completion problem," which asks how cells can accurately control the replication completion time despite the stochasticity \cite{bechhoefer07}. Here, we give full details about those calculations and go further, to investigate the idea that cells regulate the replication process in order to minimize their use of  cell ``resources" and to explore the effects of spatial regularity on the placement of origins. 

\subsection{\label{sec:review}DNA replication in eukaryotic cells}

DNA replication is a two-step process \cite{arias07}. First, potential origins --- sites where DNA synthesis may start --- are ``licensed" across the genome. For somatic cells, licensing occurs in the G1 phase of the cell cycle; for embryos, whose abbreviated cell cycles lack the G1 and G2 phases, this occurs late in the mitosis (M) phase. The process of licensing involves the formation of pre-replicative complexes (pre-RC) of proteins. Each complex is first formed through the binding of a single group of six proteins, known as the origin recognition complex (ORC), to the DNA. Each ORC, with the help of two additional proteins (Cdc6 and Cdt1), then recruits 20-40 copies of Mini Chromosome Maintenance (MCM) 2-7 hexamer rings onto the chromosome \cite{arias07}. After licensing, the second step, DNA synthesis, starts in the synthesis (S) phase. The synthesis begins with the initiation of a potential origin --- two of the MCM 2-7 rings --- triggered by the association of cyclin-dependent kinases \cite{arias07}. Once an origin is initiated, the pre-RC disassembles, and two helicases, probably the MCM 2-7 rings, move bi-directionally outward from the origin to unwind the double-stranded DNA, forming two symmetrically propagating replication forks. Polymerases are recruited behind the forks to synthesize DNA on the single-stranded DNA. When two replication forks traveling in opposite directions meet, the helicases disassemble, and the two growing strands of newly synthesized DNA are joined together by DNA ligases. This process is referred to as a \textit{coalescence}. In eukaryotic cells, the processes of origin initiation, fork progression, and domain coalescence take place at multiple sites throughout S phase until the whole genome is replicated. Re-replication is prevented because pre-RCs are ``non-recyclable" in S phase. When potential origins initiate or are passively replicated by other replication forks, pre-RCs disassemble and are inhibited from reassembling on the DNA throughout the current S phase, thereby preventing re-initiation and re-replication \cite{arias07}. 

\subsection{\label{sec:problem}The random-completion problem}

Replication in \textit{Xenopus} embryos is interesting because the process is stochastic yet the replication completion times are tightly controlled. After fertilization, a \textit{Xenopus} embryo undergoes 12 rounds of synchronous, uninterrupted, and abbreviated cell cycles (lacking G1 and G2 phases), whose durations are strictly controlled by biochemical processes that are independent of replication \cite{hyrien03, hensey97}. In contrast to the case of most somatic cells, these embryonic cells lack an efficient S/M checkpoint to delay entrance into mitosis for unusually slow replication \cite{Kimelman87}. Nonetheless, in each embryonic cell cycle, roughly 3 billion basepairs of DNA are replicated in a 20 min S phase followed by a 5 min mitosis (M) phase at 23$^{\circ}$C \cite{blow01b, cell_cycle_duration}. If replication is not completed before the end of mitosis, the cell suffers a ``mitotic catastrophe" where the chromosomes break, eventually leading to cell death \cite{prokhorova03, hensey97, hensey98}. (See Sec.~\ref{sec:in_vivo} for more discussion.) In replicating the lengthy genome, $\mathcal{O}(10^6)$ potential origins are licensed, without sequence specificity, and initiated stochastically throughout S phase \cite{hyrien93, herrick02}. One might expect that this spatiotemporal stochasticity leads to large fluctuations in replication times, which would result in frequent mitotic catastrophes. However, experiments imply that such catastrophic events for \textit{Xenopus} embryos happen only once in about 300 instances (see Sec.~\ref{sec:in_vivo}). This means that despite the stochasticity in licensing and initiations, \textit{Xenopus} embryos tightly control the duration of S phase, in order to meet the 25 min ``deadline" imposed by the cell-cycle duration.

Laskey was the first to ask whether non-sequence-specific licensing might lead to incomplete replication \cite{laskey85}. Specifically, he assumed that origins in embryonic cells initiate at the start of S phase. (This is now known not to be the case \cite{herrick02}.) He then noted that if the origins were licensed at random, they would have an exponential distribution of separations. With the estimates of the average inter-origin spacing and fork velocity known at that time, one would expect a few large gaps. The extra time needed to replicate the gaps would then imply a replication time larger than the known duration of S phase. Even though some details have changed, biologists still have such a paradox in mind when they refer to the random-completion problem \cite{blow01b}. 

Over the years, biologists have proposed two qualitative scenarios to resolve the random-completion paradox. The first scenario, the ``regular-spacing model," incorporates mechanisms that regularize the placement of potential origins despite the non-sequence specificity to suppress fluctuations in the size of inter-origin gaps \cite{walter97, blow01b}. The second scenario, the ``origin-redundancy model," uses a large excess of potential origins that are initiated with increasing probability throughout S phase to suppress the fluctuations produced by random licensing \cite{hyrien03, rhind06}. Experimentally, the observed replication kinetics in \textit{Xenopus} are compatible with the origin-redundancy model, but there is also evidence for limited regularity in the origin spacings \cite{herrick02, blow01b, lucas00}.

In this paper, we shall reformulate the random-completion problem in a more general way and investigate both scenarios using a stochastic model and Monte Carlo simulations. We consider the case in which origin initiation rates can be time dependent and non-zero throughout S phase. We then investigate how cells control the total replication time despite the non-sequence-specific placement and stochastic initiation of potential origins. As we shall see, the fluctuations in the replication times can be reduced arbitrarily if one allows an unrestricted number of initiations. As an extreme example, having an infinite number of initiations at time $t^*$ implies that replication would always finish at $t^*$. Thus, an even more general formulation of the random-completion problem is to ask how reliability in timing control can be achieved with a reasonable or ``optimal" use of resources in the cell. Of course, the terms ``reasonable", ``optimal", and ``resources" must be carefully defined.

In the following section, we review and extend the previously developed model of replication to derive the distribution of replication times \cite{jun05a, jun05b}. The results will show how replication timing can be controlled despite the stochasticity. In Sec.~\ref{sec:exp}, we use the extended model to extract replication parameters from \textit{in vivo} and \textit{in vitro} experiments. In Sec.~\ref{sec:optimization}, we compare the extracted \textit{in vivo} ``replication strategy" with the strategy that optimizes the consumption of replication forks. In Sec.~\ref{sec:LGM}, we explore the effect of spatial ordering on the replication time via a variant of the regular-spacing model. We summarize our findings in Sec.~\ref{sec:conclusion}.

\section{\label{sec:model}Modeling replication completion}

In previous work, we developed a stochastic model of DNA replication \cite{jun05a, jun05b} that was inspired by the Kolmogorov-Johnson-Mehl-Avrami (KJMA) theory of phase-change kinetics \cite{kolmogorov37, johnson39, avrami39, sekimoto84, sekimoto91, ben-naim96}. The KJMA model captures three aspects of phase transformation: nucleation of the transformed phase, growth of the nucleated domains, and coalescence of impinging domains. Making a formal analogy between phase transformations and DNA replication, we map the kinetics of the DNA replication onto a one-dimensional KJMA model with three corresponding elements: initiation of potential origins, growth of replicated domains, and coalescence of replicated domains. Note that our use of a phase-transformation model implicitly incorporates the observation that, ordinarily, re-replication is prevented.

Since we neglect any stochasticity in the movement of replication forks, the stochastic element of the model lies entirely in the placement and initiation of origins \cite{stall_fork}. The licensing and initiations can be viewed as a two-dimensional stochastic process with a spatial dimension whose range corresponds to the genome and a temporal dimension whose range corresponds to S phase. There is good evidence that the positions of the potential origins in \textit{Xenopus} embryos are almost --- but not completely --- random \cite{hyrien93, blow01b, lucas00}. In this section, we assume the spatial positions of the potential origins to be uniformly distributed across the genome for ease of calculation. We discuss the implications of origin regularity in Sec.~\ref{sec:LGM}. The temporal program of stochastic initiation times is governed by an initiation function $I(t)$, defined as the rate of initiation per unreplicated length per time. In writing down the initiation rate as a simple function of time, we are implicitly averaging over any spatial variation and neglecting correlations in neighboring initiations. The $I(t)$ deduced from a previously analyzed \textit{in vitro} experiment on \textit{Xenopus} implies that the initiation rate increases throughout S phase \cite{herrick02}. In order to explore analytically a family of initiation functions that includes such a form, we investigate the distribution of replication completion times associated with $I(t)=I_nt^n$, with $I_n$ a constant. We also examine an alternative $\delta$-function form, where all potential origins initiate at the start of S phase, as one might expect this to be the best scenario for accurate control of replication time. (In the early literature on DNA replication, biologists assumed this scenario to be true \cite{laskey85}.)

\begin{figure}
	\centering
	\includegraphics[width=3.25 in]{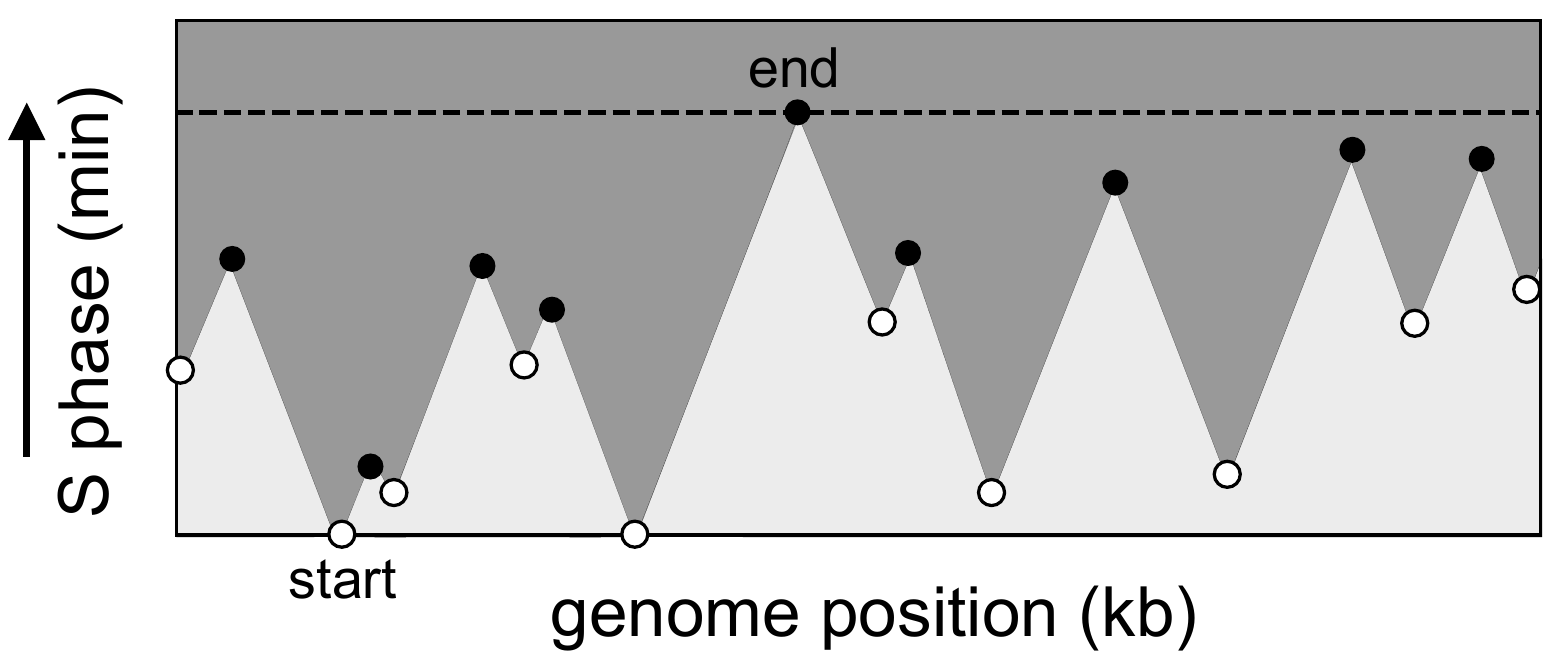} 
	\caption{Schematic of the DNA replication model. A horizontal slice in the figure represents the state of the genome at a fixed time. The lighter (darker) gray represents unreplicated (replicated) regions. Open circles denote initiated origins, while filled circles denote coalescences. The dark dotted line cuts across the last coalescence, which marks the completion of replication. The slope of the lines connecting the adjacent open and filled circles gives the inverse of the fork velocity.}
	\label{fig:rep_model}
\end{figure}

Figure~\ref{fig:rep_model} shows schematically the initiations and subsequent development of replicated domains discussed earlier. After initiation, a replicated domain grows bidirectionally outward from the origin. The growth stops when domains meet and coalescence but proceeds elsewhere. Multiple domains grow and coalesce throughout S phase until the entire genome is duplicated. We shall assume, for simplicity, that the replication fork velocity is constant. Since variations in fork velocity have been observed, a constant velocity should be interpreted as averaging over the course of S phase \cite{marheineke04, conti07}. We discuss the effect of varying fork velocities in more detail in Sec.~\ref{sec:in_vitro}.

Our model results in a deterministic growth pattern once the initiations are set. Figure~\ref{fig:rep_model} illustrates such deterministic growth and shows that, except at the edges, there is a one-to-one mapping between the initiations and the coalescences. It follows that every distribution of initiations $\phi_i(t)$ determines an associated distribution of coalescences $\phi_c(t)$. Since the completion of replication is marked by the last coalescence, the problem of determining the time needed to replicate a genome of finite length is equivalent to that of determining the distribution of times at which the last coalescence occurs. We refer to this distribution as the ``end-time" distribution $\phi_e(t)$. Below, we derive an analytical approximation to the end-time distribution function for arbitrary $I(t)$. This analytical result will allow us to investigate how licensing and initiation programs affect the timing of replication completion.

In addition to analytic results, we also carried out extensive numerical simulations of DNA replication. The simulation algorithm used is a modified version of the previously developed ``phantom-nuclei algorithm" \cite{jun05a}. The phantom-nuclei algorithm includes three main routines: the first determines the random-licensing positions and the origins' stochastic initiation times via Monte Carlo methods \cite{krauth06}; the second implements the deterministic growth; and the third eliminates passively replicated origins. Once potential origins are licensed, the algorithm can calculate the state of the genome at any time step without computing intermediate time steps. We modified our earlier code to generate end-time distributions using the bisection method to search for the first $t$ at which the replication fraction $f$ becomes 1 \cite{press07}. All programming was done using Igor Pro v. 6.01 \cite{igor}.

\subsection{\label{sec:end_time}The end-time distribution}

In previous work, we showed that for an infinitely long genome, the fraction $f$ of the genome that has replicated at time $t$ is given by \cite{jun05a}
\begin{equation}
	f(t) = 1 - e^{-2vh(t)} \; ,
 \label{eq:fraction}
\end{equation}
where $v$ is the fork velocity (assumed constant), $h(t) = \int_0^t g(t') dt'$ and $g(t) = \int_0^t I(t') dt'$.  Equation~\ref{eq:fraction} predicts that an infinite time is needed to fully duplicate the genome; however, since all real genomes are finite in length, they can be fully replicated in a finite amount of time. During the course of replication, as long as the number of replicated domains is much greater than one, the infinite-genome model is reasonably accurate. However, since the number of domains is small at the beginning and end of replication ($f\rightarrow0$ and $f\rightarrow1$), we expect discrepancies in those regimes. In particular, to calculate the finite replication time expected in a finite genome, we need to extend our previous model.

We begin by introducing the hole distribution, $n_h(x,t) = g^2(t) \exp[-g(t)x-2vh(t)]$ which describes the number of ``holes" of size $x$ per unit length at time $t$ \cite{jun05a}. A ``hole" is the biologists' term for an unreplicated domain surrounded by replicated domains. Since a coalescence corresponds to a hole of zero length, we define the coalescence distribution $\phi_c(t) \propto n_h(0,t)$. Normalizing by imposing the condition $\int_0^{\infty} \phi_c(t) dt=1$, we find
\begin{equation}
	\phi_c(t) = \frac{2vL}{N_o} g^2(t) e^{-2vh(t)} \; ,
	\label{eq:coal}
\end{equation}
where $L$ is the genome length and $N_o$ the expected total number of initiations. Note that $N_o$ is also the total number of coalescences because of the one-to-one mapping discussed in the previous subsection. One can calculate $N_o$ via
\begin{equation}
  N_o=L\int_0^{\infty}I(t)[1-f(t)]dt=L\int_0^{\infty}I(t)e^{-2vh(t)}dt \; , 
	\label{eq:number}
\end{equation} 
where the factor $[1-f(t)]$ arises because initiations can occur only in unreplicated regions. The integrand in Eq.~\ref{eq:number} divided by $N_o$ is the initiation distribution $\phi_i(t)dt$, which corresponds to the number of initiations between time $t$ and $t+dt$. 

Given the initiation distribution, we picture the initiations as sampling $N_o$ times from $\phi_i(t)$. This implies that $N_o$ independent coalescence times are sampled from $\phi_c(t)$. The replication completion time, finite on a finite genome, can then be associated with the largest value of the $N_o$ coalescence times, and the end-time distribution is the distribution of these largest values obtained from multiple sets of sampling from $\phi_c(t)$. At this point, we apply extreme-value theory (EVT) to calculate the end-time distribution. EVT is a well-established statistical theory for determining the distributional properties of the minimum and maximum values of a set of samples drawn from an underlying ``parent" distribution \cite{gumbel58, kotz00}. The properties of interest include the expected value, fluctuations around the mean, frequency of occurrence, etc. EVT plays a key role in the insurance industry, where, for example, the ``100-year-flood" problem asks for the expected maximum water level over 100 years \cite{reiss01}. In physics, EVT has attracted increasing interest and been applied to analyze crack avalanches in self-organized material \cite{caldarelli96}, degree distribution in scale-free networks \cite{moreira02}, and many other problems.

EVT is powerful because of its universality. The key theorem in EVT states that the distribution of the extremes of an independent and identically distributed random variable tends to one of three types of extreme value distributions, the Gumbel, Frechet, and Weibull distributions, depending only on the shape of the tail of the underlying distribution. The universality of the extreme value distribution with respect to the underlying distribution is similar to that of the better-known Central Limit Theorem \cite{tijms04}. For an underlying distribution with an unbounded tail that decays exponentially or faster, the distribution of the extremes tends to a Gumbel distribution. Such is the case of \textit{Xenopus} since the underlying distribution, the coalescence distribution $\phi_c(t)$, is approximately proportional to $e^{-\tau^4}$, where $\tau$ is a dimensionless time \cite{note_bilinear, note_rate_decay}. The other initiation functions we consider also lead to the Gumbel distribution.

The Gumbel distribution,
\begin{equation}
\rho(x)=\frac{1}{\beta}\exp\left(-x-e^{-x}\right),\;\;\;\;\;\;x=\frac{t-t^*}{\beta},
\label{eq:end_general}
\end{equation}
depends on only two parameters, $t^*$ and $\beta$ \cite{gumbel58,Sethna06,kotz00}. The former is a ``location" parameter that gives the mode of the distribution. The latter is a ``scale" parameter proportional to the standard deviation. We follow standard procedures to obtain $t^*$ and $\beta$ as a function of the initiation rate and the fork velocity \cite{gumbel58,Sethna06}. The main step is to recognize that the cumulative end-time distribution $\Phi_e(t)$, which has a Gumbel form, is equal to the product of $N_o$ cumulative coalescence distributions, each resulting from the same initiation distribution $\phi_i(t)$. In other words, the probability that $N_o$ coalescences occur at or before time $t$ is equivalent to the probability that the last of them occurred at or before time $t$, which is also the probability that the replication will finish at or before time $t$. For our case, we find that the mode $t^*$ is determined implicitly by 
\begin{equation}
N_o\left[1-\Phi_c(t^*)\right]=1
\label{eq:t*}
\end{equation}
and $\beta\approx1/[N_o\phi_c(t^*)]$. In Eq.~\ref{eq:t*}, $\Phi_c(t)$ is the cumulative distribution of $\phi_c(t)$; thus, $[1-\Phi_c(t)]$ is the probability that a coalescence would occur at or after time $t$. Equation~\ref{eq:t*} then implies that given a total of $N_o$ coalescences, $t^*$ is the time after which the expected number of coalescences is one, and therefore, the typical end-time. The Gumbel form of the end-time distribution is one of our main results, as it allows quantitative comparison between the fluctuations of completion times resulting from different initiation functions.

Below, we derive the end-time distribution for a power-law initiation function $I_n(t)=I_nt^n$ (where $n > -1$) and a delta-function initiation function $I_{\delta}(t)=I_{\delta}\delta(t)$. In the power-law case, $h(t)\propto t^{n+2}$, while for the $\delta$-function case, $h(t)\propto t$. From Eq.~\ref{eq:coal}, both initiation forms give rise to coalescence distributions that decay exponentially or faster, and thus, both forms will lead to an end-time distribution of the Gumbel form. Using these initiation functions, we see that the coalescence distribution given by Eq.~\ref{eq:coal} is completely determined by three parameters: the fork velocity $v$, the initiation strength given by the prefactor $I_n$ or $I_{\delta}$, and the initiation form determined by $n$ or $\delta(t)$. The relationship between these three parameters and the two Gumbel parameters reveals how different ``initiation strategies" affect the completion time.

We write the cumulative distribution $\Phi_c(t)$ of the coalescences as $1-\int_t^{\infty}\phi_c(t')dt'$. Then, using integration by parts, we obtain
\begin{equation}
 \int_t^\infty \phi_c(t')dt' = \frac{L}{N_o}g(t)e^{-2vh(t)}-\frac{L}{N_o}\int_t^\infty I(t')e^{-2vh(t')}dt' \;. 
\label{eq:2nd_term}
\end{equation}
Substituting Eq.~\ref{eq:2nd_term} into Eq.~\ref{eq:t*}, we obtain a transcendental equation
\begin{equation}
 2vh(t^*)= \ln\left[(1-\alpha)Lg(t^*)\right],\;\;\;\;
 \alpha=\frac{\int_{t^*}^\infty I(t)e^{-2vh(t)}dt}{g(t^*)e^{-2vh(t^*)}}
\label{eq:In}
\end{equation}
that relates the initiation parameters to $t^*$. For the width, Eqs.~\ref{eq:coal} and \ref{eq:In} give 
\begin{equation}
\beta=\frac{1-\alpha}{2vg(t^*)} \; ,
\label{eq:width}
\end{equation}
indicating that the width of the end-time distribution, $\beta$, is inversely proportional to $g(t^*)$, the typical number of potential origins per length. In practice, given experimentally observed quantities such as $v$, $t^*$, and $L$, we solve Eqs.~\ref{eq:In} and \ref{eq:width} numerically to determine the initiation prefactor ($I_{\delta}$ or $I_n$) and the width for different initiation forms [$\delta(t)$ or $t^n$]. Nevertheless, an analytical approximation of Eqs.~\ref{eq:In}--\ref{eq:width} is possible, as the factor $\alpha$ is often small. For instance, in the power-law $I(t)$ case, introduce a function $\eta(t)=be^{-at}$ that decays more slowly than $\phi_i(t)$. Then, imposing $\eta(t^*)=\phi_i(t^*)$ so that $\eta(t)>\phi_i(t)$ for $t>t^*$, we find $\alpha$ to be at most $\mathcal{O}(10^{-2})$. Neglecting $\alpha$, we then obtain the analytical approximations
\begin{eqnarray}
I_n&\approx&\frac{(n+1)(n+2)}{2v{t^*}^{n+2}}
\ln\left[\frac{L(n+2)}{2v{t^*}^{n+2}}\right]\\
\beta&\approx&\frac{n+1}{2vI_n{t^*}^{n+1}}
\label{eq:approx_In_width}
\end{eqnarray}
that show the explicit relationship between the initiation parameters and the Gumbel parameters.

In summary, given a realistic initiation function $I(t)$ and fork velocity $v$, we have shown that the distribution function of replication end-time tends toward a Gumbel form. We have also shown how the replication parameters relate to the location and scale Gumbel parameters analytically.

\subsection{\label{sec:control}Replication timing control}

As a first step toward understanding the solutions to the random-completion problem, we consider the end-time distributions produced by different initiation functions. From these results and the theory developed, we infer two heuristic principles for controlling the end-time distribution: the first concerns the width, while the second concerns the mode. We first explore how the width $\beta$ depends on the initiation form [$\delta(t)$ and $t^n$] by simulating the replication process while constraining the typical replication time and fork velocity to match the values inferred from \textit{in vitro} experiments: $t^*=38$ min and $v=0.6$ kb/min. (As we discuss in Sec.~\ref{sec:exp}, replication \textit{in vitro} is slower than \textit{in vivo}.) The genome length $L$ is $3.07\times10^6$ kb throughout the paper \cite{thiebaud77}. The prefactors $I_\delta$ and $I_n$ are then calculated using Eq.~\ref{eq:In}. 

\begin{figure}
	\centering
	\includegraphics[width=3.25in]{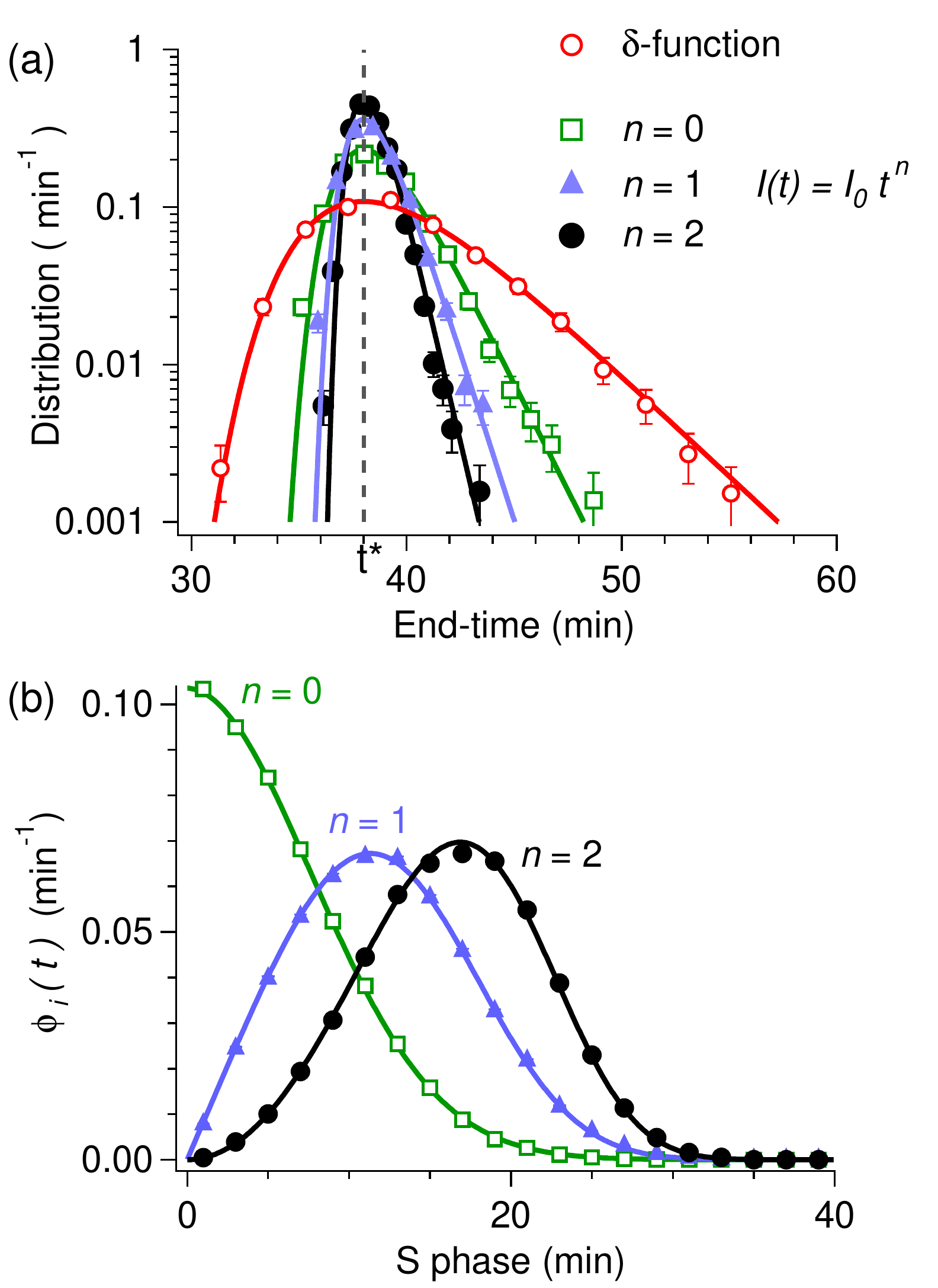} 
	\caption{(Color online) (a) The end-time distribution with fixed mode $t^*=38$ min. Markers are the results of the Monte Carlo simulations. Each distribution is estimated from 3,000 end-times. The ``$\delta$-function" corresponds to initiating all potential origins simultaneously at $t=0$ min. The $n=0$, 1, 2 cases correspond to constant, linearly increasing, and quadratically increasing initiation rates, respectively. Solid lines are Gumbel distributions with $t^*$ and $\beta$ calculated according to Eqs.~\ref{eq:In}--\ref{eq:width}. There are no fit parameters. (b) Initiation distribution $\phi_i(t)$ for $n=0$, 1, 2. Parameter values correspond to those in (a). Error bars are smaller than marker size. Solid lines are calculated from Eq.~\ref{eq:number}. Again, there are no fit parameters.}
	\label{fig:t_star}
\end{figure}

The result shown in Fig.~\ref{fig:t_star}(a) is perhaps counter-intuitive: initiating all origins in the beginning of S phase, which corresponds to a $\delta$-function $I(t)$, gives rise to the broadest distribution. Initiating origins throughout S phase narrows the end-time distribution. The narrowing is more pronounced as the power-law exponent $n$ increases. These observations can be explained by Eq.~\ref{eq:width}, which states that the width is inversely proportional to the typical density of potential origins. The physical interpretation is that having fewer potential origin sites leads to more variation in the spacing between potential origins. This in turn induces fluctuations in the largest spacings between initiated origins, which widens the end-time distribution. In this light, Fig.~\ref{fig:t_star}(a) shows that when $t^*$ is fixed, the $\delta$-function case uses the fewest potential origins and thus produces the widest distribution. In contrast, a large power-law exponent $n$ implies the use of many potential origins and thus produces a narrow distribution. In summary, the first heuristic principle is that the end-time distribution can be narrowed arbitrarily by increasing the number of potential origins in the system.

The second principle is that given an excess of potential origins, cells can initiate origins progressively throughout S phase instead of all at once, lowering the consumption of resources while still controlling the typical replication time. In S phase, initiation factors and polymerases are recyclable proteins; i.e., they can be reused once they are liberated from the DNA \cite{bell02}. Progressive initiation then allows a copy of the replicative protein to be used multiple times. Compared with initiating all origins at once, this strategy requires fewer copies of replicative proteins and thus saves resources. This notion of minimizing the required replication resources is further discussed in Sect.~\ref{sec:optimization}.

Figure~\ref{fig:t_star}(b) shows that increasing the exponent $n$ results in the ``holding back" of more and more initiations until later in S phase. Comparing this with Fig.~\ref{fig:t_star}(a), one finds that holding back initiations corresponds to narrowing the end-time distribution. Although many potential origins are passively replicated and thus never initiate, the timing of replication can still be accurately controlled, as initiations now occur in the ``needed places." Since the probability of initiation inside a hole is proportional to the size of the hole, the held-back initiations are more likely to occur in large holes. This filling mechanism is made efficient by increasing $I(t)$ toward the end of S phase so that any remaining large holes are increasingly likely to be covered.

One subtle point of the origin-redundancy scenario is that although the potential origins are licensed at random, the spacings between initiated origins form a distribution $\rho_i(s)$ with a non-zero mode that contrasts with the exponential distribution of spacings between potential origins. An example of the $\rho_i(s)$ is shown later in Sec.~\ref{sec:LGM}. In earlier literature, before experiments showed that initiations can take place throughout S phase, biologists believed that all potential origins initiate at the start of S phase. In this $\delta$-function case, the distribution of the inter-potential-origin spacing is the same as that of the spacing between fired origins (inter-origin spacing). However, with an increasing $I(t)$, a peak will arise in $\rho_i(s)$ because closely spaced potential origins are not likely to all initiate but be passively replicated by a nearby initiation. This passive replication effect suppresses the likelihood of having small inter-origin spacings and thus creates a non-zero mode in the spacing distribution. One should be careful not to confuse the two distributions.

In conclusion, we have shown that a large excess of potential origins suppresses fluctuations in the size of inter-potential-origin gaps while the strategy of holding back initiations allows control of the typical replication time. In the next section, we review what is known experimentally about DNA replication in \textit{Xenopus} embryos, in light of the analysis we have just presented.

\section{\label{sec:exp} Analysis of replication experiments}

In the previous section, we showed that given an initiation function and a fork velocity, one can find the associated end-time distribution using EVT. In this section, we review what is known experimentally about these quantities in \textit{Xenopus} embryos. There have been two classes of experiments: \textit{in vivo}, where limited work has been done \cite{prokhorova03, hensey97, hensey98}, and \textit{in vitro}, where rather more detailed studies have been performed on cell-free extracts \cite{herrick00, marheineke01, blow01b, herrick02}. Typically, embryo replication \textit{in vivo} takes about 20 minutes of the (abbreviated) 25-minute cell cycle \cite{hyrien03, cell_cycle_duration}. As we discuss below, \textit{in vivo} experiments imply that replication ``failure" --- incomplete replication by the end of the cell cycle --- is very unlikely, occurring only once in about 300 instances. The \textit{in vitro} experiments on cell-free extracts give more detailed information about the replication process, including an estimate of the \textit{in vitro} initiation function $I_{vitro}(t)$. However, the typical replication time \textit{in vitro} is about 38 min, not 20 min, and it is not obvious how one can apply the results learned from the \textit{in vitro} experiments to the living system. Below, we propose a way to transform $I_{vitro}(t)$ into an estimate of the \textit{in vivo} initiation function $I_{vivo}(t)$ that satisfies the failure probability of the \textit{in vivo} system.

\subsection{\label{sec:in_vivo} The \textit{in vivo} experiments}

A low replication-failure rate is remarkable because \textit{Xenopus} embryos lack an efficient S/M checkpoint to delay cell cycle progression when replication is incomplete \cite{hyrien03}. If chromosomes separate before replication is complete, cells suffer ``mitotic catastrophe," which leads to apoptosis \cite{prokhorova03}. Thus, a low failure rate in embryonic cells implies that replication timing is precisely controlled by the initiation function and fork velocity. Mathematically, we can test whether an initiation function is realistic by calculating the rate of mitotic catastrophe $F$ it implies. To evaluate $F$, we first choose a time $t^{**}$ at which mitotic catastrophe occurs if replication is not fully completed. Then,
\begin{equation}
F \equiv \int_{t^{**}}^\infty \phi_e(t)dt \;= \;1-\Phi_e(t^{**}) \; .
\label{eq:constraint}
\end{equation}

As a first step in estimating $F$, we identify $t^{**}$ with the cell cycle time ($\approx$ 25 min) \cite{cell_cycle_duration}. Our identification is justified by observations that imply that replication can continue throughout mitosis, if needed \cite{prokhorova03}. Thus, even if the bulk of replication is completed before entering mitosis, small parts of the genome may continue to replicate, essentially until the cell totally divides. However, if while the cell is dividing, unreplicated regions of the chromosome segregate, mitotic catastrophe would cause the two daughter cells to inherit fragmented chromosomes. 

\begin{figure}
	\centering
	\includegraphics[width=3.25 in]{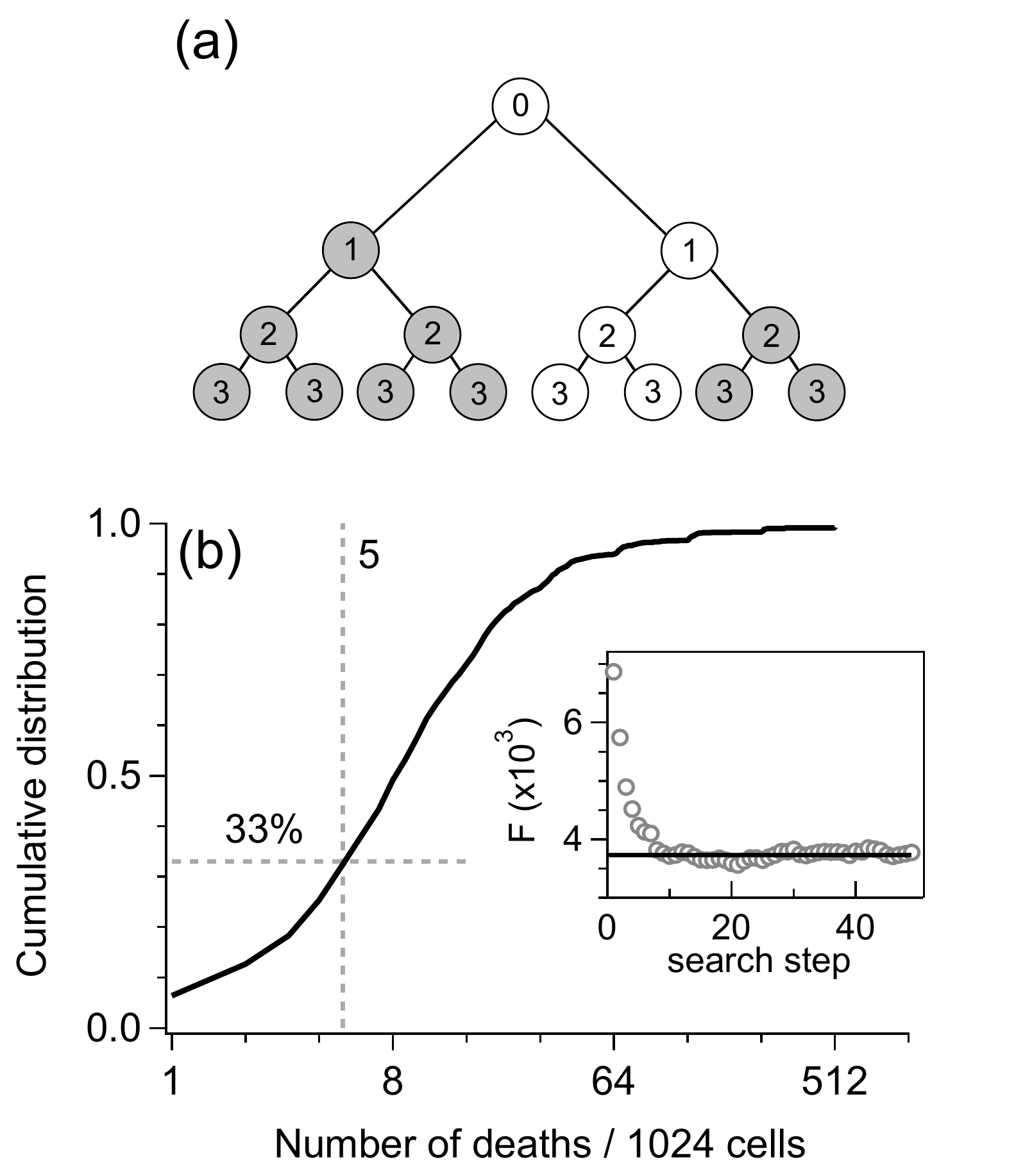} 
	\caption{(a) Schematic diagram of the simple model described in the text. Open circles represent normal proliferating cells, while filled circles are abnormal cells. The numbers indicate the round of cleavage. Once a cell fails to replicate properly, all its progeny will be abnormal. (b) Cumulative distribution of the number of dead cells at gastrulation (between cleavage 10 and 11) generated using Monte Carlo simulation. The distribution satisfies the constraint that 33\% of the embryos have 5 or fewer abnormal cells. The inset shows the convergence of the gradient search to $F=3.73\pm0.01 \times 10^{-3}$. The average and standard deviation of the mean are computed over the last 40 values.}
	\label{fig:mp}
\end{figure}

Having identified $t^{**}$, we estimate $F$ using data from an experiment on DNA damage in embryos \cite{hensey97, hensey98}. In \cite{hensey97}, Hensey and Gautier found that cells with massive DNA damage (induced by radiation) will continue to divide through 10 generations. Then, at the onset of gastrulation, which occurs between the $10^{th}$ and $11^{th}$ cleavages, an embryo triggers a developmental checkpoint that activates programmed cell death. The role of cell death is to eliminate abnormal cells before entering the next phase of development, where the the embryo's morphology is constructed via cell migration. In Hensey and Gauthier's study, abnormal cells were detected using TUNEL staining, a technique for detecting DNA fragmentation in cells. In a later work investigating the spatial-temporal distribution of cell deaths in \textit{Xenopus} embryos, they reported that, at gastrulation, 67\% of 237 embryos, each containing $1024$ cells, had more than 5 TUNEL-stained cells \cite{hensey98}. We can estimate $F$ from the above observations using a simple model based on the following four elements:
\begin{enumerate}
	\item All cells divide; each produces two cells.
	\item If a cell has an abnormal chromosome, all its progeny are abnormal because replication can at best duplicate the parent's chromosome.
	\item Failure to replicate all DNA before the end of a cell cycle is the main cause of abnormal chromosomes and leads to apoptosis at gastrulation.
	\item All normal cells in all rounds of cleavage have the same probability $F$ of becoming abnormal because of incomplete replication.
\end{enumerate}
A schematic depiction of our model is shown in Fig.~\ref{fig:mp}(a).

The above model can be described by a standard Galton-Watson (GW) branching process \cite{kimmel02}, where the number of proliferating progeny generated by a normal cell is an independent and identically-distributed random variable. GW processes obey recursion relations that can be solved analytically using probability generating functions; however, the solution in our case is too complex to be helpful. We thus turned to numerical analysis. 

We used Monte Carlo methods to simulate the branching process outlined above. Each embryo, after going through $10$ rounds of division, contains $m$ abnormal cells that commit apoptosis before the $11^{th}$ division. Simulating $N$ embryos results in a distribution of number of deaths. We then compare the evaluation of the cumulative distribution at 5 death events with the reported likelihood, which states that 33\% of the time, there are 5 or fewer dead cells in 1024 cells \cite{hensey98}. Figure~\ref{fig:mp}(b) shows the cumulative distribution that matches the reported numbers. To find $F$, we used a gradient-based method for finding roots of stochastic functions. In this case, the input is the failure rate $F$, and the function evaluates the number and likelihood of deaths via a Monte Carlo simulation of the branching process of 237 embryos. We found that the numbers reported in \cite{hensey98} imply $F=3.73\pm0.01$ [Fig.~\ref{fig:mp}(b) inset] \cite{note_low_F, spall03}. In summary, we inferred the failure rate in \textit{Xenopus} embryo replication to be about 1 in 300.

Comparing Eq.~\ref{eq:constraint} with the standard cumulative Gumbel distribution given by the integral of Eq.~\ref{eq:end_general}, one can relate the quantities $t^{**}$ and $F$ to the Gumbel parameters via
\begin{equation}
t^{**}= t^*-\beta(t^*) \ln\left[\ln \left(\frac{1}{1-F}\right)\right] \;.
\label{eq:t**tot*}
\end{equation}
For $F \ll 1$, the expression simplifies to $t^{**} \approx t^* - \beta(t^*)\ln(F)$, which implies that the end-time is insensitive to the exact value of $F$: an order-of-magnitude estimate suffices.

\subsection{\label{sec:in_vitro}Connecting \textit{in vitro} to \textit{in vivo}}

As discussed above, the most detailed experiments on replication in \textit{Xenopus} have been conducted on cell-free egg extracts. In previous work \cite{herrick02}, we modeled a molecular-combing experiment on such an \textit{in vitro} system and inferred the time-dependent initiation function $I_{vitro}(t)$ (approximately quadratic \cite{note_bilinear, note_rate_decay}), a fork velocity of 0.6 kb/min (averaged over S phase \cite{marheineke04}), and a typical replication time $t^*$ of 38 min. In contrast, the typical replication time in living embryos is only 20 min. While it is generally believed that DNA replication in the two settings occurs in a similar way, the overall duration of S phase is an obvious difference that must be reconciled. We thus have a dilemma: the known replication parameters, $v$ and $I(t)$, are extracted from \textit{in vitro} experiments while the failure rate $F$ is derived from observations of cells \textit{in vivo}. Is it possible to ``transpose" the results from the \textit{in vitro} experiments to the \textit{in vivo} setting? Although any such transformation is obviously speculative, we propose here a simple way that is consistent with known experimental results.

We hypothesize that, except for the fork velocity, replication is unaltered between the \textit{in vitro} and \textit{in vivo} systems. The subtlety is that there are several conceivable interpretations of ``unaltered" replication. One could keep $I_{vitro}(t)$ the same; however, this is not reasonable in that the dramatic increase in $I_{vitro}(t)$, at $t\approx 17.4$ min, would be moved from the midpoint of replication to the end \cite{herrick02}. Alternatively, one could express the initiation function in terms of the fraction of replication, ie. $I=I(f)$, and preserve this function. In this case, one would need a fork velocity of about 2.2 kb/min to produce the extracted \textit{in vivo} failure rate. Although this is a reasonable fork speed in systems such as the \textit{Drosophila} embryo, it is about twice the maximum fork speed observed in \textit{Xenopus} embryonic replication \textit{in vitro} \cite{marheineke04}. The third possibility is to preserve the maximum number of simultaneously active replication forks. Intuitively, this is plausible as each replication fork implies the existence of a large set of associated proteins. The maximum fork density then gives the minimum number of copies of each protein set required. Thus, we are in effect assuming that the numbers of replicative proteins remains the same in both cases.

\begin{figure}
	\centering
	\includegraphics[width=3.25 in]{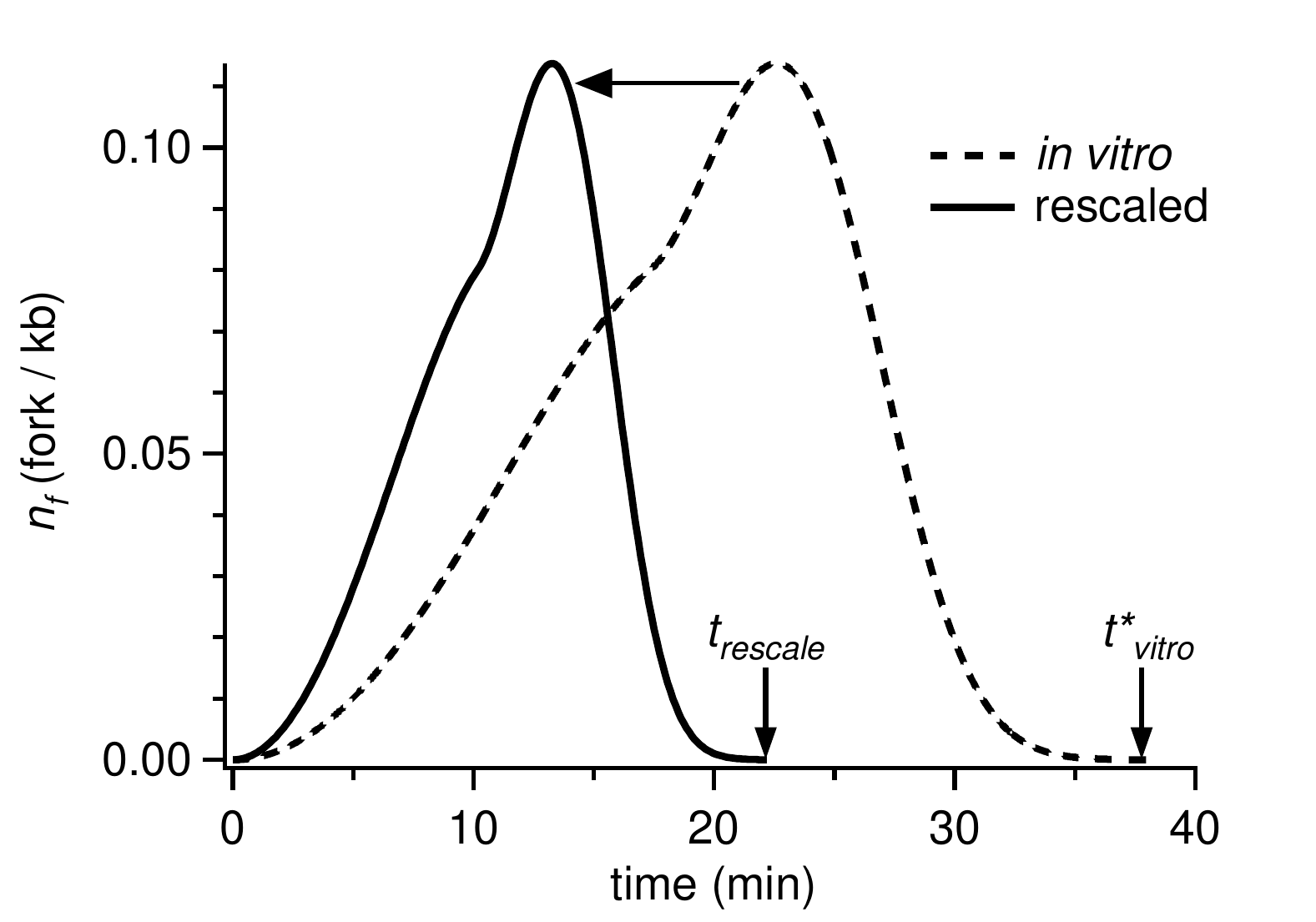} 
	\caption{Density of simultaneously active replication forks throughout S phase, $n_f(t)$. The dotted curve corresponds to the \textit{in vitro} fork usage while the solid curve is the rescaled fork usage that satisfies the constraints $t^{**}=25$ min and $F=0.00373$. The rescaled $n_f(t)$ is generated using $I_{vivo}(t/t^*_{vivo})\approx 2I_{vitro}(t/t^*_{vitro})$ and $v=1.030$ kb/min.}
	\label{fig:trans}
\end{figure}

The simplest way to preserve fork usage is to rescale the density of forks active at time $t$,
\begin{equation}
 n_f(t)=\frac{1}{2v}\frac{df}{dt}=g(t)e^{-2vh(t)}\;,
\label{eq:fork_t}
\end{equation}
linearly in time so that
\begin{equation}
n_f^{vivo}\left(\frac{t}{t_{scale}}\right)= n_f^{vitro}\left(\frac{t}{t^*_{vitro}}\right) \;,
\label{eq:fork}
\end{equation}
where $t^*_{vitro} \approx 38$ min and $t_{scale}$ is chosen so that $t^{**}=25$ min and $F=3.73\pm0.01 \times 10^{-3}$. We found that the \textit{in vitro} fork usage is preserved by using the rescaling $I_{vivo}(t/t_{scale})\approx 2 I_{vitro}(t/t^*_{vitro})$ and $v=1.030\pm0.001$ kb/min (Fig.~\ref{fig:trans}). The error on $v$ is the consequence of the uncertainty in $F$. This velocity has a significant interpretation. In a recent experiment, Marheineke and Hyrien found that the fork velocity \textit{in vitro} is not constant but decreases linearly from about 1.1 kb/min to 0.3 kb/min at the end of S phase \cite{marheineke04}. The decrease in fork velocity suggests that \textit{in vitro} replication progressively depletes rate-limiting factors (e.g., dNTP) throughout S phase. We suggest that our extracted $v\approx1$ kb/min means that \textit{in vivo} systems are able to maintain the concentration of rate-limiting factors, perhaps by regulating their transport across the nuclear membrane \cite{moore99, nguyen00}, to maintain a roughly constant fork velocity throughout S phase. In summary, by preserving the rescaled version of the \textit{in vitro} fork usage rate, we have transformed $I_{vitro}(t)$ into an $I_{vivo}(t)$ that results in reasonable replication parameters and reproduces the \textit{in vivo} failure rate.

\section{\label{sec:optimization}Optimizing fork activity}

\begin{figure}
	\centering
	\includegraphics[width=3.25in]{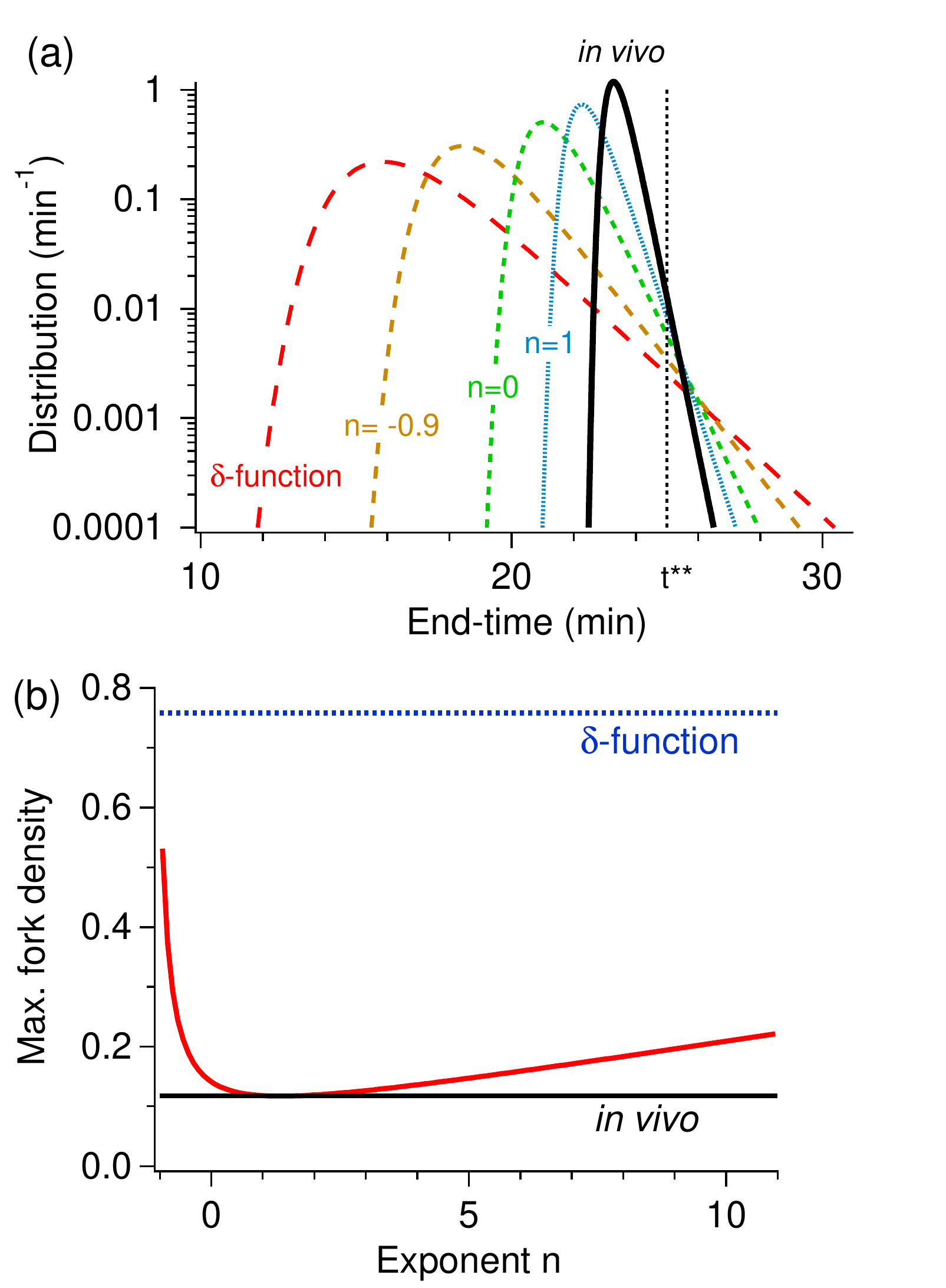} 
	\caption{(Color online) (a) Replication end-time distribution with $t^{**}$ fixed to be 25 min and $F=0.00373$. Similar to Fig.~\ref{fig:t_star}(a), the width decreases with an increase in the exponent $n$. (b) Typical maximum number of simultaneously active forks. The curve is obtained from extracting the maximum value of $n_f(t)$ for different exponents $n$.}
	\label{fig:t**}
\end{figure}

The random-completion problem mentioned in Sec.~\ref{sec:intro} can be quantitatively recast into a problem of searching for an initiation function that produces the \textit{in vivo} failure rate constraint in Eq.~\ref{eq:constraint}. In Fig.~\ref{fig:t**}(a), we show that any initiation form with the proper prefactor can satisfy the constraint on the integral of the end-time distribution, including the transformed \textit{in vivo} initiation function. Can we then understand why \textit{Xenopus} embryos adopt the roughly quadratic $I(t)$ and not some other function of time? 

To explore this question, we calculate for the different cases of $I(t)$ the maximum number of simultaneously active forks. Figure~\ref{fig:t**}(b) shows that initiating all origins at the start of S phase [setting $I(t)\sim\delta(t)$] requires a higher maximum than a modestly increasing $I(t)$. At the other extreme, a too rapidly increasing $I(t)$ (high exponent $n$) also requires many copies of replicative machinery because the bulk of replication is delayed and needs many forks close to the end of S phase to finish the replication on time. Thus, intuitively, one expects that an intermediate $I(t)$ that increases throughout S phase --- but not too much --- would minimize the use of replicative proteins. Figure~\ref{fig:t**}(b) hints that the \textit{in vivo} initiation function derived from \textit{in vitro} experiments may be close to such an optimal $I(t)$, as the number of resources required by $I_{vivo}(t)$ is close to the minimum of the power-law case.

It is not immediately clear which replication resources should be optimized. Here, we propose that the maximum number of simultaneously active forks be minimized. Above, we argued that the maximum of $n_f(t)$ gives the minimum number of copies of the proteins required for DNA synthesis. Moreover, since the unwinding and synthesis of DNA at the forks create torsional stress on the chromosomes, minimizing the number of active forks would minimize the complexity of the chromosome topology, which may help maintain replication fidelity \cite{postow01}. For these reasons, the maximum number of active forks is a plausible limiting factor that causes replication to proceed the way it does. Below, we calculate the optimal $I(t)$ and compare it with $I_{vivo}(t)$.

\begin{figure}[!ht]
	\includegraphics[width=3.25in]{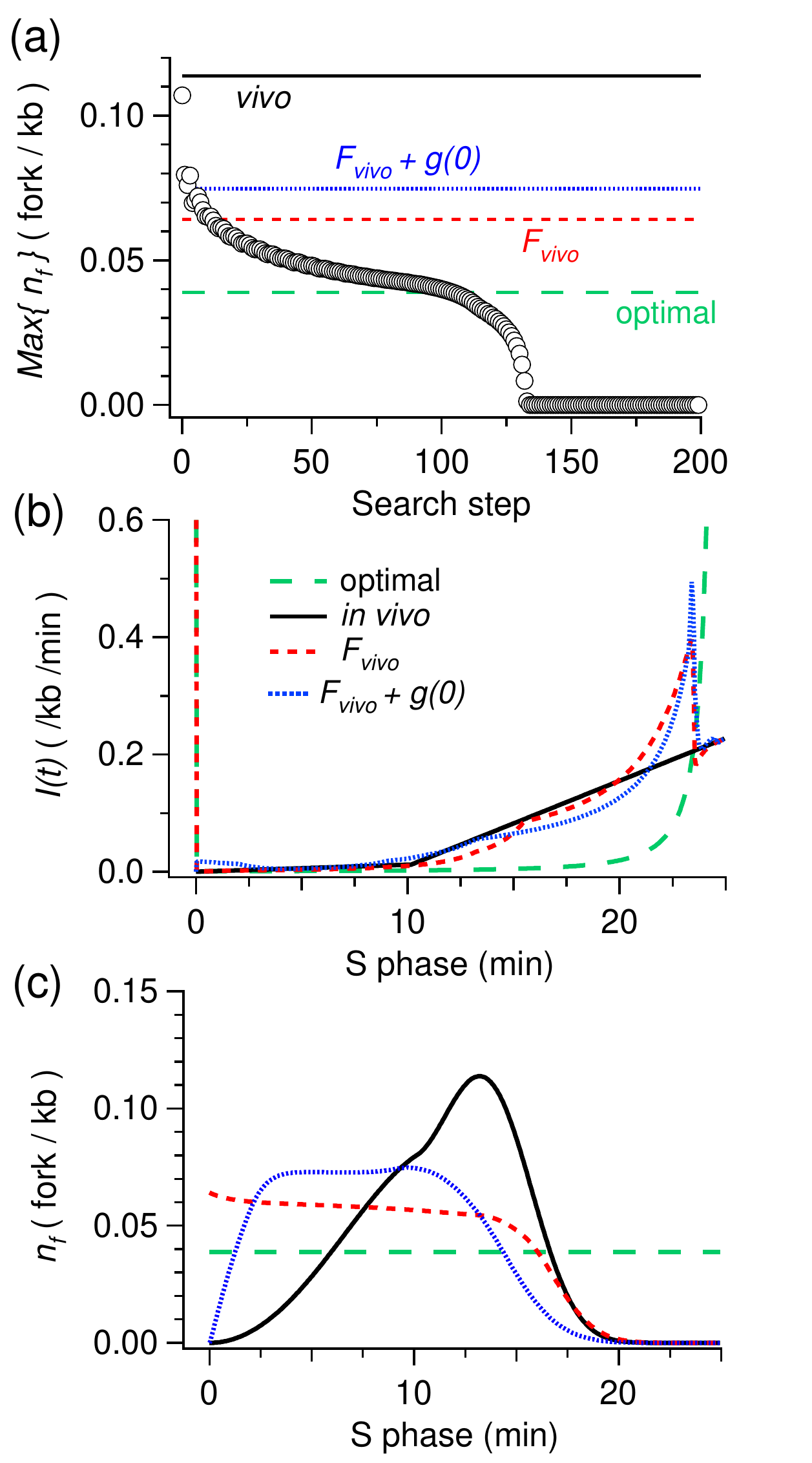} 
	\caption{(Color online) Results of a numerical search for optimal initiation functions under various constraints. The label ``\textit{vivo}" corresponds to the \textit{in vivo} case; ``optimal" corresponds to optimizing maximum fork density with no constraint (corresponds to Eq.~\ref{eq:i_opt_fork}); ``$F_{vivo}$" corresponds to optimization with the constraint that the failure rate be equal the $F_{vivo}$ extracted in Sec.~\ref{sec:in_vivo}; ``$F_{vivo}+g(0)$" corresponds to optimization with the constraint of $F_{vivo}$ and the constraint that $g(0)=0$. (a) Finite difference stochastic approximation search. The markers shows the search for the case of minimizing the $max\{n_f\}$ with no constraint and no boundary condition. The horizontal lines are the maximum fork density for different search conditions. (b) Initiation rate $I(t)$. The $I_{vivo}$ shown is in the bi-linear form \cite{note_bilinear}. (c) Fork density $n_f(t)$. Line types correspond to the same cases as in (b).}
	\label{fig:opti}
\end{figure}

The number of forks active at time $t$ is given by $n_f(t)=2g(t)\exp[-2vh(t)]$. One can find the $I(t)$ that optimizes the maximum of $n_f(t)$ by minimizing 
\begin{equation}
	n_{max}[I(t)] = \lim_{p \to \infty}
	\left[ \int_0^{t^{**}} n_f\left[I(t)\right]^p \; dt \right]^{1/p} \; .
\label{eq:opt_fork}
\end{equation}
This is a common analytic method to optimize the maximum of a function \cite{skogestad05}. The trick is to analytically calculate the Euler-Lagrange equations for finite $p$ and then take the limit $p \rightarrow \infty$, where the contribution of the maximum dominates the integrand. The associated Euler-Lagrange equation is
\begin{equation}
	\ddot{h}(t) = 2v \dot{h}^2(t) \; ,
	\label{eq:euler}
\end{equation}
where we recall that $\ddot{h}(t) = I(t)$ and $\dot{h}(t) = g(t)$.  Note that Eq.~\ref{eq:euler} is independent of $p$, suggesting that the optimal $n_f(t)$ does not have a peak. Solving Eq.~\ref{eq:euler} subject to the boundary condition that the replication fraction be 0 at $t=0$ [ie., $h(0) = 0$] and 1 at $t=t^{**}$, we obtain
\begin{equation}
	I_{opt}(t) = \frac{1}{2vt^{**}} \left[ \delta(t) + \frac{1}{t^{**}} \frac{1}{(1 - t/t^{**})^2} \right] \;  .
	\label{eq:i_opt_fork}
\end{equation}
Inserting the result from Eq.~\ref{eq:i_opt_fork} into Eq.~\ref{eq:fork_t}, one sees that $n_f(t) = 1/vt^{**}$ indeed is constant throughout S phase and is about three times smaller than the maximum number of simultaneously active forks \textit{in vivo} [Fig.~\ref{fig:opti}(c)]. This optimal solution, like $I_{vivo}(t)$, increases slowly at first, then grows rapidly toward the end of S phase [Fig.~\ref{fig:opti}(b)]. However, this initiation function is unphysical, as the diverging initiation probability at $t\rightarrow t^{**}$ implies an infinite number of initiations at the end of S phase. In effect, a constant fork density implies that when the protein complexes associated with two coalescing forks are liberated, they instantly find and attach to unreplicated parts of the chromosome. It also implies that at the end of S phase, all the replication forks would be active on a vanishingly small length of unreplicated genome. Both implications are unrealistic.

To find a more realistic solution, we tamed the behavior of the initiation rate for $t\rightarrow t^{**}$ by adding a constraint. A natural constraint to impose is that the failure rate \textit{in vivo} be satisfied. The infinite initiations at $t=t^{**}$ implied by Eq.~\ref{eq:i_opt_fork} means that the replication always finishes exactly at $t^{**}$ and the failure rate is zero. Therefore, having a non-zero failure rate would force the number of initiations to be finite. This constraint is also consistent with the idea that the replication process is shaped by the evolutionary pressure of survival. The new optimization quantity is then
\begin{equation}
	J\left[I(t)\right] = max\left\{n_f\left[I(t)\right]\right\} + \lambda \left\{F\left[I(t)\right]-F_{vivo}\right\},
\label{eq:opt_fork_mod}
\end{equation}
where the first term is the maximum of the fork density, and the second term is a penalty function that increases $J$ for $F\neq F_{vivo}$. The strength of the penalty is set by the Lagrange multiplier $\lambda$. The time associated with $F$ is $t^{**}=25$ min throughout this section.

Substituting Eq.~\ref{eq:opt_fork} into the first term of Eq.~\ref{eq:opt_fork_mod} and applying the method of variational calculus, we obtained an integro-differential equation that turns out to be stiff mathematically and thus difficult to solve. The difficulty in analytic methods is that the gradient of Eq.~\ref{eq:opt_fork} is highly nonlinear and that $F$ depends on $t^*$, which is not readily expressible in terms of the basic replication parameters $I(t)$ and $v$. For these reasons, we turned to a gradient-free numerical method called finite difference stochastic approximation (FDSA) \cite{spall03}. Although this search method is used for stochastic functions (as the name suggests), the method is just as suitable for deterministic functions. The basic concept is that the gradient of a function, which encodes the steepest-decent direction toward a local minimum, can be approximated by a finite difference of the function. The advantage of this method is that we can replace the complicated evaluation of the variation $\delta J[I(t)]$ by the easily calculable difference $J[I+\Delta I]-J[I-\Delta I]$. 

Figure~\ref{fig:opti} shows the results of the FDSA search. We performed FDSA under several different conditions, with the initial search function being $I_{vivo}(t)$. First, we investigate the case where the optimization objective $J$ is simply $max\{n_f\}$, with no further constraint or boundary condition [except $n_f(t)>0$]. The markers in Fig.~\ref{fig:opti}(a) shows that the optimal solution lingers near $max\{n_f\}=0.05$ (slow decrease in $J$) and then goes to the global minimum (zero). In the transient regime (search step between 50 to 100), the fork density evolves from a bell curve to a constant, which is the form of the calculated optimal solution. For search step $>$ 100, the fork density (a constant) decreases to zero if no constraint is imposed. This zero solution corresponds to the case where no initiation or replication occurs. However, when the boundary condition used in the calculation (replication finished at $t^{**}$) is imposed, the FDSA algorithm indeed finds the $n_f(t)=1/vt^{**}$ optimal solution (data not shown).

The second search was implemented following Eq.~\ref{eq:opt_fork_mod}, where the constraint in $F$ is added. Figure~\ref{fig:opti}(c) shows that the fork solution is no longer a constant because the tail needs to decrease to satisfy $F=F_{vivo}$. The corresponding effect on the $I(t)$ is a decrease toward the end of S phase [Fig.~\ref{fig:opti}(b)]. Interestingly, the mechanism of spreading out the fork density to minimize the maximum fork usage seen in the analytical calculation is still present here, as shown by the plateau at the beginning. The $I(t)$ then behaves as predicted by Eq.~\ref{eq:i_opt_fork} for most of S phase --- a $\delta$-function at the beginning followed by a rate that increases sharply at the end of S phase.

In the third search, in addition to Eq.~\ref{eq:opt_fork_mod}, we imposed that there be no burst of initiation at the beginning of S phase [$g(0)=0$], as seen in experiments. Figure~\ref{fig:opti}(c) shows that with the addition of each constraint, the maximum of the fork density increases toward the \textit{in vivo} value. Further, besides satisfying the constraints and boundary conditions, the fork density profiles show a common feature of forming as lengthy a plateau as possible to minimize the maximum. The resulting $I(t)$ is very similar to $I_{vivo}$ [Fig.~\ref{fig:opti}(b)]. 

However, there are still some differences between the result of the third search and $n^{vivo}_f$. In particular, the optimal fork solution increases much faster at the beginning of S phase than $n^{vivo}_f$ does, to spread out the fork activities. Minimizing the maximum number of initiations also leads to the same feature of a fast initial increase in the initiation activities followed by a plateau. These observations then suggest that while minimizing the maximum of simultaneously active replicative proteins may be a factor that determines the replication pattern, there must a stronger limiting factor at the beginning of S phase. A plausible hypothesis is that the copy number of at least some of the replicative proteins is small to start with but gradually increases with nuclear import throughout S phase \cite{hyrien08_private}. With this additional constraint, the replication activities at the beginning of S phase are limited by the small numbers of available replicative proteins. In conclusion, the optimization method presented in this section is useful because it connects the replication process with an optimal criterion that is plausibly connected with evolutionary selection pressure. This allows one to explore the limiting factors of replication. Moreover, the method is general in that it allows one to explore the consequences of a wide range of possible constraints.

\section{\label{sec:LGM}The Lattice-genome model: from random to periodic licensing}

Until now, we have assumed a spatially random distribution of potential origins. In this section, we explore the implications of spatial ordering among the potential origins on the end-time distribution. We have two motivations. First, an ``obvious" method for obtaining a narrow end-time distribution is to space the potential origins periodically and initiate them all at once. However, such an arrangement would not be robust, as the failure of just one origin to initiate would double the replication time. Still, the situation is less clear if initiations are spread out in time, as the role of spatial regularity in controlling inter-origin spacing would be blurred by the temporal randomness.

Our second motivation is that there is experimental evidence that origins are not positioned completely at random. A completely random positioning implies that the distribution of gaps between potential origins is exponential, resulting in many small inter-potential-origin spacings. However, in an experiment of plasmid replication in \textit{Xenopus} egg extracts, Lucas \textit{et al.} found no inter-origin gap smaller than 2 kb \cite{lucas00}. In a previous analysis, we also observed that, assuming random licensing, one expects more inter-origin gaps less than 8 kb than were observed and fewer between 8--16 kb \cite{jun05b}. Second, experiments have suggested a qualitative tendency for origins to fire in groups, or clusters \cite{blow01b}. These findings collectively imply that there is some spatial regularity in the \textit{Xenopus} system, perhaps through a ``lateral inhibition" of licensing potential origins too closely together. Our goal is to find an ``ordering threshold," at which point the resulting end-time distribution starts to deviate from the random-licensing case. We will then argue that one needs not worry about effects of regular spacing in the \textit{Xenopus} system because the experimental degree of ordering is well below this threshold.

To investigate spatial ordering, we change the continuous genome to a ``lattice genome" with variable lattice spacing $d_l$. Potential origins can be licensed only on the lattice sites. For $d_l \rightarrow 0$, the lattice genome becomes continuous, and the model recovers the random-licensing case. As $d_l$ increases, the lattice genome has fewer available sites for licensing potential origins, and the fraction of licensed sites increases. In this scenario, the spacing between initiated origins take on discrete values --- multiples of $d_l$. One can imagine that a further increase in $d_l$ would eventually lead to a critical $d_l$, where every lattice site would have a potential origin. This scenario corresponds to an array of periodically licensed origins, which leads to a periodic array of initiated origins with spacing $d_l$. Thus, by increasing a single parameter $d_l$, we can continuously interpolate from complete randomness to perfect periodicity. 

In order to compare regularized licensing to random licensing, we impose that the cell environment, including the concentrations of replicative proteins such as CDK, helicases, and polymerases, be the same in spite of varying $d_l$. This constraint implies that while the potential origins may be distributed along the genome differently, the total initiation probability across the genome is conserved. We then write
\begin{equation}
I(x,t)=d_l\,I(t)\sum_{n=0}^{L/d_l} \delta(x-nd_l)\; ,
\label{eq:I_cl}
\end{equation}
where $x$ is the position along the genome. This equation shows that as the number of lattice sites $L/d_l$ is reduced via an increase in $d_l$, the initiation probability for each site is enhanced, resulting in more efficient potential origins. This implies a tradeoff between the ``quantity" and ``efficiency" of potential origins. 

\begin{figure}
	\includegraphics[width=3.25in]{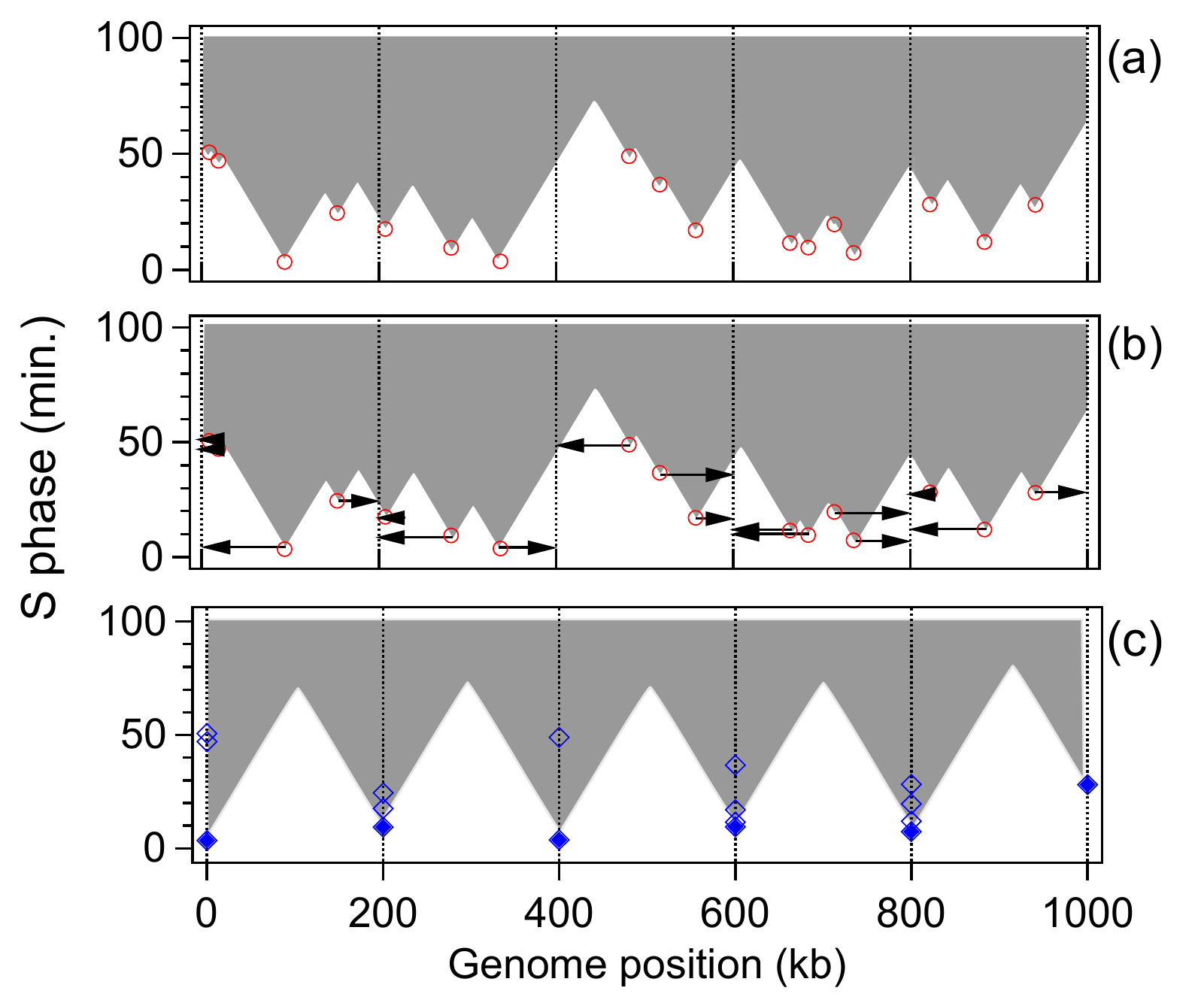} 
	\caption{ Schematic diagram of licensing on a lattice genome. (a) Realization of replication using random-licensing ($d_l=0$ case). The gray (white) area represents replicated (unreplicated) domains. Circles denote initiations. (b) Origins are forced to their nearest lattice sites (marked by vertical lines at multiples of $d_l=200$ kb), while initiation times remain the same. (c) The result of the shift in origin positions. Open markers represent ``phantom origins" that do not contribute to the replication; filled markers denote the actual origins. Going from $d_l=0$ kb in (a) to 200 kb in (c), the average initiation time decreased from about 22 min to about 10 min.}
	\label{fig:con_lattice}
\end{figure}

Figure~\ref{fig:con_lattice} illustrates this concept of tradeoff, showing how Eq.~\ref{eq:I_cl} connects random licensing to ordered licensing. A realization of random licensing is shown in Fig.~\ref{fig:con_lattice}(a). Since Eq.~\ref{eq:I_cl} modifies only the spatial distribution of origins relative to our previous $I(t)$, the effect of going from a continuous genome to a lattice genome is equivalent to shifting the randomly licensed origins to their nearest lattice sites while preserving their initiation times [Fig.~\ref{fig:con_lattice}(b)]. In doing so, we obtained Fig.~\ref{fig:con_lattice}(c), which shows multiple initiations on a lattice site. Since re-initiation is forbidden in normal replication, on each site only the earliest initiation contributes to the replication. The later initiations are ``phantom origins" that illustrate how ordering reduces the number of initiations but enhances the efficiency of potential origin sites. The increase in efficiency is indicated by the decrease in the average initiation times between the two scenarios.

\begin{figure}
	\includegraphics[width=3.25in]{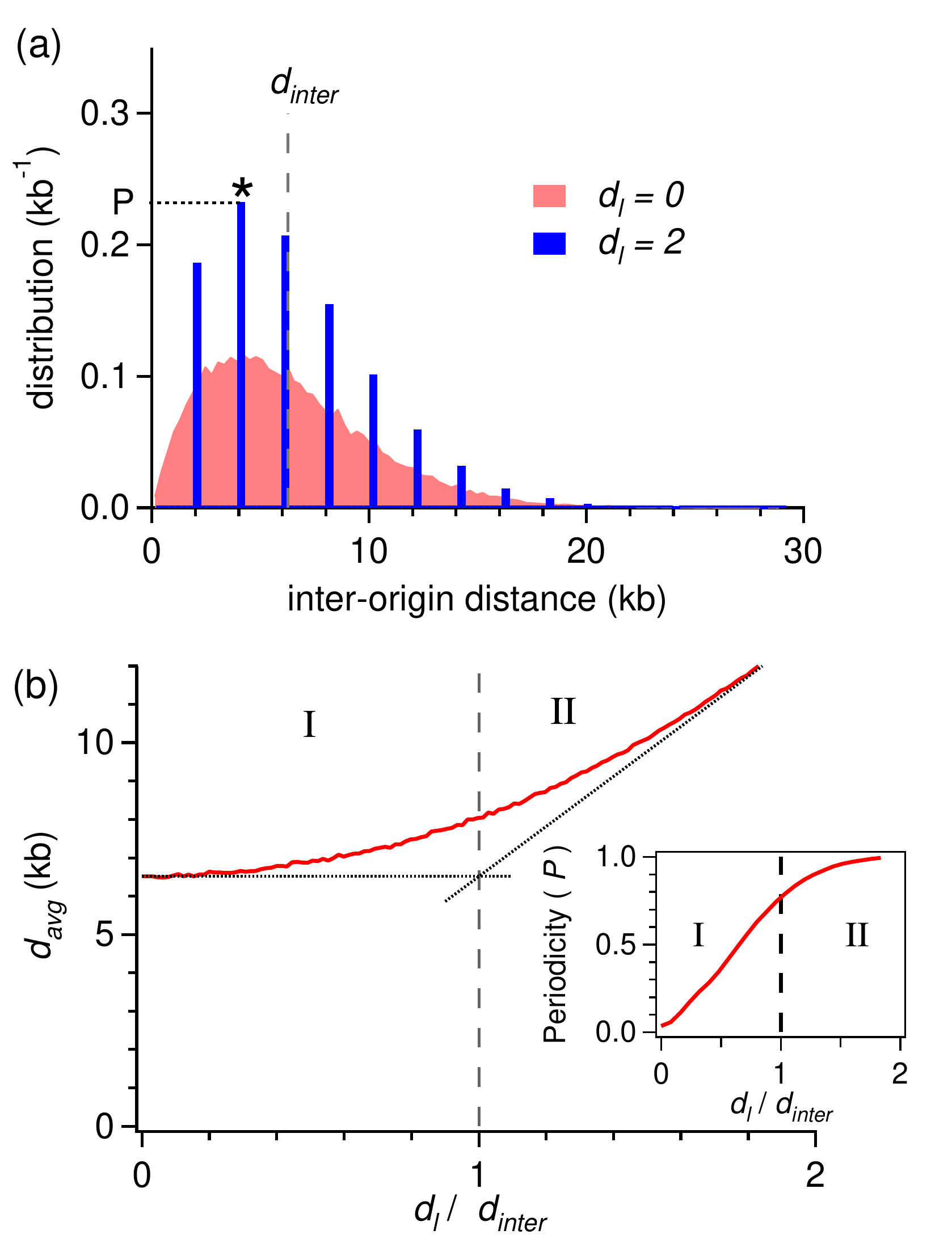} 
	\caption{(a) The distribution of spacings between initiated origins, $\rho_i(s)$, for the $d_l=0$ and 2 kb cases (2 kb is chosen to mimic the minimal spacing between origins reported in \cite{lucas00}). The initiation rate and fork velocity are those obtained in Sec.~\ref{sec:in_vitro}. The mean of the continuous distribution ($d_l=0$ kb case) is marked $d_{inter}$ and is $\approx$ 6.5 kb. The mode of the discrete distribution ($d_l=2$ kb case) is marked by ``$\;\star\;$." The probability $P$ at the mode (0.23 in this case) is defined to be the periodicity, a measure of ordering in the system. (b) Average inter-origin spacing $d_{avg}$ as a function of $d_l$. There is a gradual transition from Regime I to Regime II. In Regime I ($d_l \leq d_{inter}$), $d_{avg}$ is asymptotically independent of $d_l$ for $d_l \rightarrow 0$. In Regime II ($d_l>d_{inter}$), $d_{avg}$ is asymptotically linearly proportional to $d_l$. Inset shows the periodicity $P$ as a function of $d_l$.}
	\label{fig:inter_ori}
\end{figure}

Having outlined the rules for licensing, we now introduce two quantities, ``periodicity" $P$ and $d_{inter}$, that will be useful in later discussions of how $d_l$ alters the end-time distribution. We first look at $\rho_i(s)$, the distribution of the spacing between initiated origins, where $s$ is the inter-origin spacing. Figure~\ref{fig:inter_ori}(a) shows two $\rho_i(s)$'s: the continuous one corresponds to random licensing, while the discrete one corresponds to setting $d_l$ to 2 kb. The two distributions are different because of the discretization effect of the lattice genome: origins can have separations that are only multiples of $d_l$. As $d_l$ increases, one expects a dominant spacing to appear in the system. We characterize this ordering effect by defining the periodicity $P$, the probability at the mode of the discrete inter-origin-spacing distribution. As an example, the $d_l=2$ kb distribution shown in Fig.~\ref{fig:inter_ori}(a) has $P=0.23$, indicating that 23\% of the spacings have the same value. In the fully periodic case, the probability at the mode is 1, as all the spacings have the same value: the system is then 100\% periodic ($P=1$). For $d_l\rightarrow0$, $P$ should be interpreted as the mode of $\rho_i(s)$ times a vanishingly small $\Delta s$ ($\sim d_l$). Thus, $P\rightarrow0$ in the small $\Delta s$ limit, as there will be no inter-origin spacings sharing the same size.

In interpolating from random licensing to periodic licensing, one expects that the average inter-origin spacing $d_{avg}$ would change from being $d_l$-independent to being linearly dependent on $d_l$. Indeed, from Fig.~\ref{fig:inter_ori}(b), which shows $d_{avg}$ as a function of $d_l$, we can label two asymptotes and thereby identify two regimes. We first introduce $d_{inter}$ to be the average inter-origin spacing of the $d_l=0$ kb case. For $d_l\rightarrow0$, we see that $d_{avg}$ asymptotically approaches $d_{inter}$. In contrast, for large $d_l$ (when all lattice sites are occupied), we see $d_{avg}$ approaching the asymptote $d_{avg}=d_l$. The intersection of the two asymptotes is precisely at $d_l=d_{inter}$. We therefore identify two regimes with Regime I being $d_l \leq d_{inter}$ and Regime II being $d_l > d_{inter}$. Physically, the weak $d_l$ dependence in Regime I suggests that the system is spatially random, whereas the asymptotically linear behavior in Regime II indicates that the system is becoming periodic. 

The length scale $d_{inter}$ encodes the two factors that determine the distribution of inter-origin spacings. The first factor is the passive replication of closely positioned potential origins, which suppresses the likelihood of having small inter-origin spacings. The second factor is based on the low probability of randomly licensing two far-away origins, which reduces the probability of having large inter-origin gaps. Both of these effects can be seen in Fig.~\ref{fig:inter_ori}(a). 

When $d_l$ exceeds $d_{inter}$, the typical spacing between potential origins ($\sim d_l$) exceeds the typical range of passive replication and approaches the typical largest spacing of the random-licensing case. This means that potential origins are not likely to be passively replicated or positioned farther than $d_l$ apart (note that the next smallest spacing $2d_l$ is quite large). The inset in Fig.~\ref{fig:inter_ori}(b), which shows the periodicity $P$ as a function of $d_l$, strengthens this notion that for $d_l>d_{inter}$, the system enters a nearly periodic regime where $P$ has saturated.

\begin{figure}
	\includegraphics[width=3.25in]{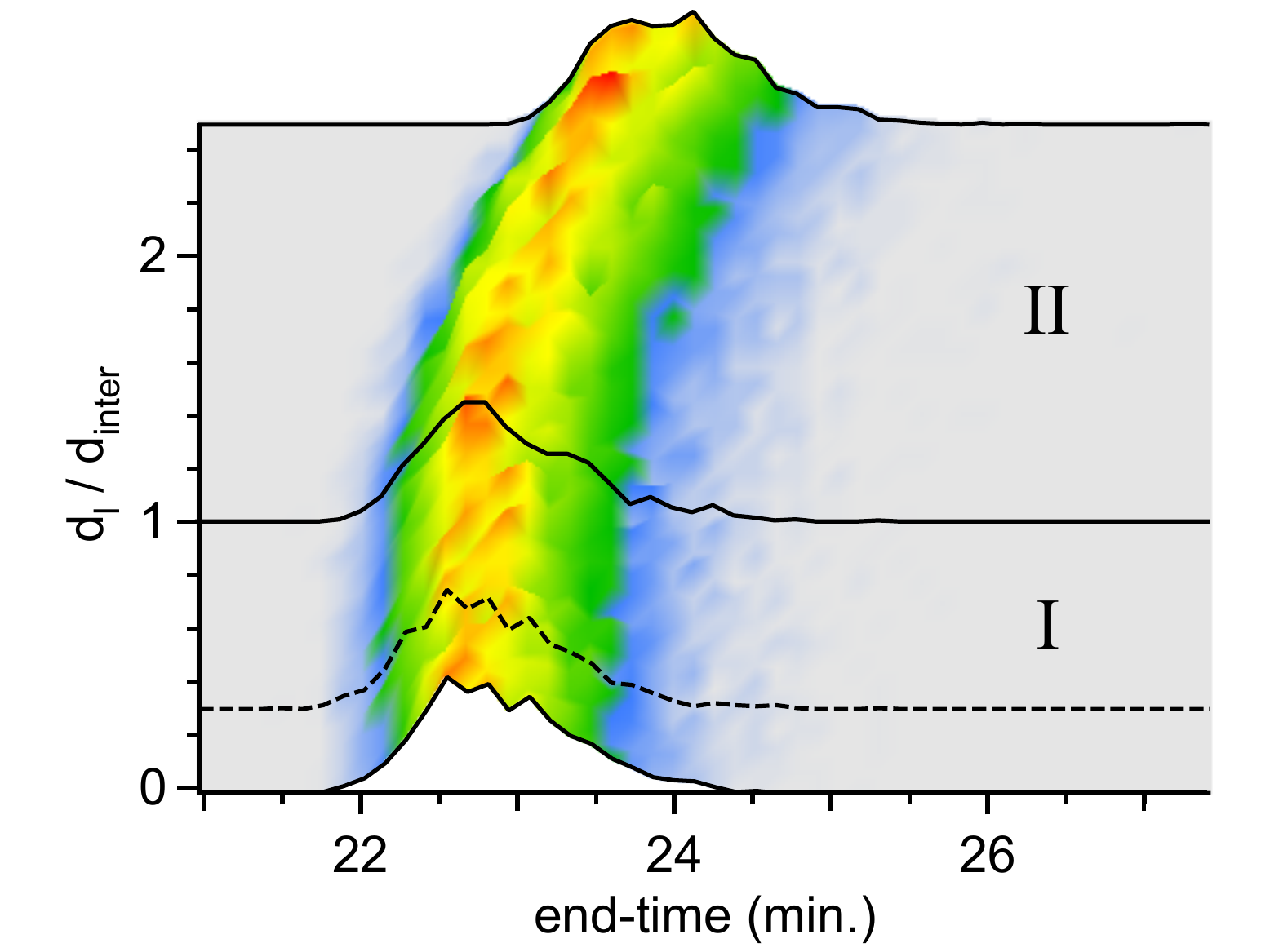} 
	\caption{(Color online) Evolution of the end-time distribution with increasing spatial ordering due to increasing $d_l$. Each horizontal profile is an end-time distribution. In Regime I, the end-time distribution does not change appreciably; in Regime II, the mode shifts to the right. The ordering threshold is at $d_l=d_{inter}\approx 6.5$ kb. The dashed line shows the $d_l=2$ kb end-time distribution, which corresponds to the lateral inhibition ordering observed experimentally \cite{lucas00}.}
	\label{fig:reg}
\end{figure}

Our main result is Fig.~\ref{fig:reg}, which shows how the end-time distribution changes with increasing $d_l$. The initiation function used in the simulation is the power-law approximation of the $I_{vivo}(t)$ found in Sect.~\ref{sec:in_vitro}, transformed using Eq.~\ref{eq:I_cl}. The fork velocity and failure rate used are as extracted in Sect.~\ref{sec:exp}. There are again two distinct regimes separated by the ordering threshold $d_{inter} \approx 6.5$ kb. Below the threshold (Regime I), the end-time distribution is nearly independent of $d_l$. Above the threshold (Regime II), the mode shifts to the right. The width is unaltered.

To understand the changes in going from Regime I to Regime II, we note that in Eq.~\ref{eq:t*}, $t^*$ depends on the number of initiations $N_o$. On average, $N_o$ is unaffected when the number of lattice sites available is in excess ($\frac{N_o}{L/d_l} > 1$). This means that $t^*$ starts to change only when $d_l=L/N_o$ which is precisely $d_{inter}$. In Regime II, the minimum time to replicate the smallest gap between potential origins, $d_l/v$, becomes significant compared to the temporal randomness resulting from stochastic initiation. In effect, $t^*\approx d_l/v+t_{avg}$, where $t_{avg}$ is the average initiation time. We tested numerically that the mean and standard deviation of the initiation times both decrease sigmoidally, for $d_l/d_{inter}>3$. Thus, for the range of $d_l$ shown in Fig.~\ref{fig:reg}, one expects $t^*\propto d_l$ in Regime II, while the width should be unaltered. 

In \textit{Xenopus} embryos, the inhibition zone observed in plasmid replication corresponds to $d_l \approx 2$ kb. This is shown as the dashed line in Fig.~\ref{fig:reg}. The value is well below the ordering threshold of $d_{inter}\approx 6.5$ kb, strongly suggesting that the experimentally observed spatial ordering is not strong enough to be relevant to the random-completion problem in embryonic replication. Recall that biologists have proposed two models for solving the random-completion problem, the regular-spacing model and the origin-redundancy model. Our results suggest that to employ the regular-spacing strategy, a cell would need mechanisms for measuring the spacing between origins and also mechanisms that ensure the early initiation of each potential origin. Such mechanisms have not been found to date. On the other hand, the cell can solve the problem by using the origin-redundancy strategy, where more potential origins are licensed than needed. Cells can then accurately control the replication time by increasing the initiation rate, perhaps by importing the required proteins into the cell nucleus. Both mechanisms in the latter strategy are ``open loop" in that they do not require any information about the replication state of the cell.

\section{\label{sec:conclusion}Conclusion}

In this paper, we have extended the stochastic nucleation-and-growth model of DNA replication to describe not only the kinetics of the bulk of replication but also the statistics of replication quantities at the end of replication. Using the model, we have quantitatively addressed the random-completion problem, which asks how stochastic licensing and initiation lead to the tight control of replication end-times observed in systems such as \textit{Xenopus} embryos. We found that the fluctuation of the end-times can be suppressed by increasing the number of potential origins, while the typical end-time can be adjusted by holding back initiations until later in S phase. Further, we analyzed the effect of spatial ordering on the replication end-time using a lattice-genome. Our results show that the end-time distribution is not affected until an ordering threshold is reached. We also showed that the observed ordering effect of lateral inhibition in \textit{Xenopus} is well below the ordering threshold. 

Comparing the two proposed solutions to the random-completion problem, the origin-redundancy model and the regular-spacing model, our results strongly suggest that embryonic replication takes the origin-redundancy approach. In fact, given that initiations are stochastic in time, the regular-spacing model does not improve the end-time control much even when potential origins are periodically positioned. This suggests that our stochastic model describes embryonic replication well and that spatial ordering mechanisms play a minor role in regulating replication times.

We have also found the optimal $I(t)$ that minimizes the maximum number of simultaneously active forks. Like the observed \textit{in vitro} initiation function, it increases throughout S phase except for the end. Further pursuit of the optimization problem with more detailed model may reveal the rate-limiting factors in replication, which have not been identified to date. Further, an open issue not addressed by our model is the observation that there is a weak clustering effect in the initiations of neighboring origins \cite{blow01b}. To model this effect, one can introduce correlations in licensing, initiation, and fork progression based on localization of replication foci \cite{meister07}, chromatin structures \cite{jun04}, or some other mechanisms. We do not expect that correlations will modify the scenario we have presented here significantly, as the most significant effect of correlations, an increase in spatial ordering, would not be important even at exclusion-zone sizes that are much larger than observed (e.g., 10 kb).

Among the various cases of replication programs, replication in bacteria is the most well understood --- DNA synthesis starts at a single, sequence-specific genome site and proceeds to completion \cite{lodish08}. With this case, the genome-wide regulation of the replication process is deterministic and strictly governed by biochemical effects. In this work, we modeled a very different type of replication program, where both the licensing and initiation timings are strongly influenced by stochastic effects. This type of stochastic replication strategy is usually present in embryos, especially those that develop outside of parents' body, for rapid development of the embryos. We showed how stochastic effects ensure the fast and reliable replication needed for rapid development.

In between these two special cases lie all other replication programs, where the licensing mechanisms are more complicated than simple sequence targeting or a Poisson process \cite{gilbert01}. These include replication in non-embryonic (somatic) cells \cite{anglana03}, in simple organisms such as yeast \cite{patel06}, and in cancer cells where abnormal replication such as re-replication can occur \cite{micco06}. For instance, replication experiments done on fission-yeast cells also show an excess number of potential origins. In that case, the positions of the origins are associated with a sequence asymmetry of the genome instead of being completely sequence independent \cite{heichinger06}. The experiments also show that different origins have different efficiencies, suggesting that initiation timings are not entirely stochastic. In future work, we shall modify our model to study such systems, which may lead to a fuller understanding of how replication is regulated by the genome organization and by DNA replication strategies that are the outcome of evolution.

\begin{acknowledgments}

We thank Nick Rhind and Michel Gauthier for helpful discussions and a thorough reading of the manuscript. We thank Olivier Hyrien for informing us of unpublished results and data. This work was funded by NSERC (Canada) and by HFSP.

\end{acknowledgments}


\begin{thebibliography}{63}
\expandafter\ifx\csname natexlab\endcsname\relax\def\natexlab#1{#1}\fi
\expandafter\ifx\csname bibnamefont\endcsname\relax
  \def\bibnamefont#1{#1}\fi
\expandafter\ifx\csname bibfnamefont\endcsname\relax
  \def\bibfnamefont#1{#1}\fi
\expandafter\ifx\csname citenamefont\endcsname\relax
  \def\citenamefont#1{#1}\fi
\expandafter\ifx\csname url\endcsname\relax
  \def\url#1{\texttt{#1}}\fi
\expandafter\ifx\csname urlprefix\endcsname\relax\def\urlprefix{URL }\fi
\providecommand{\bibinfo}[2]{#2}
\providecommand{\eprint}[2][]{\url{#2}}

\bibitem[{\citenamefont{Herrick and Bensimon}(2008)}]{herrick08}
\bibinfo{author}{\bibfnamefont{J.}~\bibnamefont{Herrick}} \bibnamefont{and}
  \bibinfo{author}{\bibfnamefont{A.}~\bibnamefont{Bensimon}},
  \bibinfo{journal}{Chromosoma} \textbf{\bibinfo{volume}{117}},
  \bibinfo{pages}{243–} (\bibinfo{year}{2008}).

\bibitem[{\citenamefont{Diffley}(1996)}]{diffley96}
\bibinfo{author}{\bibfnamefont{J.~F.} \bibnamefont{Diffley}},
  \bibinfo{journal}{Genes \& Dev.} \textbf{\bibinfo{volume}{10}},
  \bibinfo{pages}{2819} (\bibinfo{year}{1996}).

\bibitem[{\citenamefont{Arias and Walter}(2007)}]{arias07}
\bibinfo{author}{\bibfnamefont{E.~E.} \bibnamefont{Arias}} \bibnamefont{and}
  \bibinfo{author}{\bibfnamefont{J.~C.} \bibnamefont{Walter}},
  \bibinfo{journal}{Genes \& Dev.} \textbf{\bibinfo{volume}{21}},
  \bibinfo{pages}{497–} (\bibinfo{year}{2007}).

\bibitem[{\citenamefont{Hensey and Gautier}(1997)}]{hensey97}
\bibinfo{author}{\bibfnamefont{C.}~\bibnamefont{Hensey}} \bibnamefont{and}
  \bibinfo{author}{\bibfnamefont{J.}~\bibnamefont{Gautier}},
  \bibinfo{journal}{Mech. Dev.} \textbf{\bibinfo{volume}{69}},
  \bibinfo{pages}{183} (\bibinfo{year}{1997}).

\bibitem[{\citenamefont{Micco et~al.}(2006)\citenamefont{Micco, Fumagalli,
  Cicalese, Piccinin, Gasparini, Luise, Schurra, Garre, Nuciforo, Bensimon
  et~al.}}]{micco06}
\bibinfo{author}{\bibfnamefont{R.~D.} \bibnamefont{Micco}},
  \bibinfo{author}{\bibfnamefont{M.}~\bibnamefont{Fumagalli}},
  \bibinfo{author}{\bibfnamefont{A.}~\bibnamefont{Cicalese}},
  \bibinfo{author}{\bibfnamefont{S.}~\bibnamefont{Piccinin}},
  \bibinfo{author}{\bibfnamefont{P.}~\bibnamefont{Gasparini}},
  \bibinfo{author}{\bibfnamefont{C.}~\bibnamefont{Luise}},
  \bibinfo{author}{\bibfnamefont{C.}~\bibnamefont{Schurra}},
  \bibinfo{author}{\bibfnamefont{M.}~\bibnamefont{Garre}},
  \bibinfo{author}{\bibfnamefont{P.~G.} \bibnamefont{Nuciforo}},
  \bibinfo{author}{\bibfnamefont{A.}~\bibnamefont{Bensimon}},
  \bibnamefont{et~al.}, \bibinfo{journal}{Nature}
  \textbf{\bibinfo{volume}{444}}, \bibinfo{pages}{638} (\bibinfo{year}{2006}).

\bibitem[{\citenamefont{Sancar et~al.}(2004)\citenamefont{Sancar,
  Lindsey-Boltz, \"{U}nsal Ka\c{c}maz, and Linn}}]{sancar04}
\bibinfo{author}{\bibfnamefont{A.}~\bibnamefont{Sancar}},
  \bibinfo{author}{\bibfnamefont{L.~A.} \bibnamefont{Lindsey-Boltz}},
  \bibinfo{author}{\bibfnamefont{K.}~\bibnamefont{\"{U}nsal Ka\c{c}maz}},
  \bibnamefont{and} \bibinfo{author}{\bibfnamefont{S.}~\bibnamefont{Linn}},
  \bibinfo{journal}{Annu. Rev. Biochem.} \textbf{\bibinfo{volume}{73}},
  \bibinfo{pages}{39} (\bibinfo{year}{2004}).

\bibitem[{\citenamefont{Bensimon et~al.}(1994)\citenamefont{Bensimon, Simon,
  Chiffaudel, Croquette, Heslot, and Bensimon}}]{bensimon94}
\bibinfo{author}{\bibfnamefont{A.}~\bibnamefont{Bensimon}},
  \bibinfo{author}{\bibfnamefont{A.}~\bibnamefont{Simon}},
  \bibinfo{author}{\bibfnamefont{A.}~\bibnamefont{Chiffaudel}},
  \bibinfo{author}{\bibfnamefont{V.}~\bibnamefont{Croquette}},
  \bibinfo{author}{\bibfnamefont{F.}~\bibnamefont{Heslot}}, \bibnamefont{and}
  \bibinfo{author}{\bibfnamefont{D.}~\bibnamefont{Bensimon}},
  \bibinfo{journal}{Science} \textbf{\bibinfo{volume}{265}},
  \bibinfo{pages}{2096} (\bibinfo{year}{1994}).

\bibitem[{\citenamefont{Herrick and Bensimon}(1999)}]{herrick99}
\bibinfo{author}{\bibfnamefont{J.}~\bibnamefont{Herrick}} \bibnamefont{and}
  \bibinfo{author}{\bibfnamefont{A.}~\bibnamefont{Bensimon}},
  \bibinfo{journal}{Biochimie} \textbf{\bibinfo{volume}{81}},
  \bibinfo{pages}{859} (\bibinfo{year}{1999}).

\bibitem[{\citenamefont{Herrick et~al.}(2000)\citenamefont{Herrick,
  Stanislawski, Hyrien, and Bensimon}}]{herrick00}
\bibinfo{author}{\bibfnamefont{J.}~\bibnamefont{Herrick}},
  \bibinfo{author}{\bibfnamefont{P.}~\bibnamefont{Stanislawski}},
  \bibinfo{author}{\bibfnamefont{O.}~\bibnamefont{Hyrien}}, \bibnamefont{and}
  \bibinfo{author}{\bibfnamefont{A.}~\bibnamefont{Bensimon}},
  \bibinfo{journal}{J. Mol. Biol.} \textbf{\bibinfo{volume}{300}},
  \bibinfo{pages}{1133} (\bibinfo{year}{2000}).

\bibitem[{\citenamefont{Marheineke and Hyrien}(2001)}]{marheineke01}
\bibinfo{author}{\bibfnamefont{K.}~\bibnamefont{Marheineke}} \bibnamefont{and}
  \bibinfo{author}{\bibfnamefont{O.}~\bibnamefont{Hyrien}},
  \bibinfo{journal}{J. Biol. Chem.} \textbf{\bibinfo{volume}{276}},
  \bibinfo{pages}{17092} (\bibinfo{year}{2001}).

\bibitem[{\citenamefont{Herrick et~al.}(2002)\citenamefont{Herrick, Jun,
  Bechhoefer, and Bensimon}}]{herrick02}
\bibinfo{author}{\bibfnamefont{J.}~\bibnamefont{Herrick}},
  \bibinfo{author}{\bibfnamefont{S.}~\bibnamefont{Jun}},
  \bibinfo{author}{\bibfnamefont{J.}~\bibnamefont{Bechhoefer}},
  \bibnamefont{and} \bibinfo{author}{\bibfnamefont{A.}~\bibnamefont{Bensimon}},
  \bibinfo{journal}{J. Mol. Biol.} \textbf{\bibinfo{volume}{320}},
  \bibinfo{pages}{741} (\bibinfo{year}{2002}).

\bibitem[{\citenamefont{Hyrien and M\'echali}(1993)}]{hyrien93}
\bibinfo{author}{\bibfnamefont{O.}~\bibnamefont{Hyrien}} \bibnamefont{and}
  \bibinfo{author}{\bibfnamefont{M.}~\bibnamefont{M\'echali}},
  \bibinfo{journal}{EMBO J.} \textbf{\bibinfo{volume}{12}},
  \bibinfo{pages}{4511} (\bibinfo{year}{1993}).

\bibitem[{\citenamefont{Jun et~al.}(2005)\citenamefont{Jun, Zhang, and
  Bechhoefer}}]{jun05a}
\bibinfo{author}{\bibfnamefont{S.}~\bibnamefont{Jun}},
  \bibinfo{author}{\bibfnamefont{H.}~\bibnamefont{Zhang}}, \bibnamefont{and}
  \bibinfo{author}{\bibfnamefont{J.}~\bibnamefont{Bechhoefer}},
  \bibinfo{journal}{Phys. Rev. E} \textbf{\bibinfo{volume}{71}},
  \bibinfo{pages}{011908} (\bibinfo{year}{2005}).

\bibitem[{\citenamefont{Jun and Bechhoefer}(2005)}]{jun05b}
\bibinfo{author}{\bibfnamefont{S.}~\bibnamefont{Jun}} \bibnamefont{and}
  \bibinfo{author}{\bibfnamefont{J.}~\bibnamefont{Bechhoefer}},
  \bibinfo{journal}{Phys. Rev. E} \textbf{\bibinfo{volume}{71}},
  \bibinfo{pages}{011909} (\bibinfo{year}{2005}).

\bibitem[{\citenamefont{Bechhoefer and Marshall}(2007)}]{bechhoefer07}
\bibinfo{author}{\bibfnamefont{J.}~\bibnamefont{Bechhoefer}} \bibnamefont{and}
  \bibinfo{author}{\bibfnamefont{B.}~\bibnamefont{Marshall}},
  \bibinfo{journal}{Phys. Rev. Lett.} \textbf{\bibinfo{volume}{98}},
  \bibinfo{pages}{098105} (\bibinfo{year}{2007}).

\bibitem[{\citenamefont{Hyrien et~al.}(2003)\citenamefont{Hyrien, Marheineke,
  and Goldar}}]{hyrien03}
\bibinfo{author}{\bibfnamefont{O.}~\bibnamefont{Hyrien}},
  \bibinfo{author}{\bibfnamefont{K.}~\bibnamefont{Marheineke}},
  \bibnamefont{and} \bibinfo{author}{\bibfnamefont{A.}~\bibnamefont{Goldar}},
  \bibinfo{journal}{BioEssays} \textbf{\bibinfo{volume}{25}},
  \bibinfo{pages}{116} (\bibinfo{year}{2003}).

\bibitem[{\citenamefont{Kimelman et~al.}(1987)\citenamefont{Kimelman,
  Kirschner, and Scherson}}]{Kimelman87}
\bibinfo{author}{\bibfnamefont{D.}~\bibnamefont{Kimelman}},
  \bibinfo{author}{\bibfnamefont{M.}~\bibnamefont{Kirschner}},
  \bibnamefont{and} \bibinfo{author}{\bibfnamefont{T.}~\bibnamefont{Scherson}},
  \bibinfo{journal}{Cell} \textbf{\bibinfo{volume}{48}}, \bibinfo{pages}{399}
  (\bibinfo{year}{1987}).

\bibitem[{\citenamefont{Blow et~al.}(2001)\citenamefont{Blow, Gillespie,
  Francis, and Jackson}}]{blow01b}
\bibinfo{author}{\bibfnamefont{J.~J.} \bibnamefont{Blow}},
  \bibinfo{author}{\bibfnamefont{P.~J.} \bibnamefont{Gillespie}},
  \bibinfo{author}{\bibfnamefont{D.}~\bibnamefont{Francis}}, \bibnamefont{and}
  \bibinfo{author}{\bibfnamefont{D.~A.} \bibnamefont{Jackson}},
  \bibinfo{journal}{J. Cell Biol.} \textbf{\bibinfo{volume}{152}},
  \bibinfo{pages}{15} (\bibinfo{year}{2001}).

\bibitem[{cel()}]{cell_cycle_duration}
\bibinfo{note}{The durations of the embryonic cell cycle depends on
  temperature. For this paper, we take the cell cycle time to be 25 min at
  23$^{\circ}$C \cite{hyrien08_private}. A typical duration of S phase used is
  $\approx$ 20 min \cite{blow01b, hyrien03}. Longer times (25 min for S phase
  and 30 min for the cell cycle) have been observed at 20$^{\circ}$C
  \cite{hyrien08_private}.}

\bibitem[{\citenamefont{Prokhorova et~al.}(2003)\citenamefont{Prokhorova,
  Mowrer, Gilbert, and Walter}}]{prokhorova03}
\bibinfo{author}{\bibfnamefont{T.~A.} \bibnamefont{Prokhorova}},
  \bibinfo{author}{\bibfnamefont{K.}~\bibnamefont{Mowrer}},
  \bibinfo{author}{\bibfnamefont{C.~H.} \bibnamefont{Gilbert}},
  \bibnamefont{and} \bibinfo{author}{\bibfnamefont{J.~C.}
  \bibnamefont{Walter}}, \bibinfo{journal}{Proc. Natl. Acad. Sci.}
  \textbf{\bibinfo{volume}{100}}, \bibinfo{pages}{13241}
  (\bibinfo{year}{2003}).

\bibitem[{\citenamefont{Hensey and Gautier}(1998)}]{hensey98}
\bibinfo{author}{\bibfnamefont{C.}~\bibnamefont{Hensey}} \bibnamefont{and}
  \bibinfo{author}{\bibfnamefont{J.}~\bibnamefont{Gautier}},
  \bibinfo{journal}{Dev. Biol.} \textbf{\bibinfo{volume}{203}},
  \bibinfo{pages}{36} (\bibinfo{year}{1998}).

\bibitem[{\citenamefont{Laskey}(1985)}]{laskey85}
\bibinfo{author}{\bibfnamefont{R.~A.} \bibnamefont{Laskey}},
  \bibinfo{journal}{J. Embryol. Expe. Morphol. Suppl.}
  \textbf{\bibinfo{volume}{89}}, \bibinfo{pages}{285} (\bibinfo{year}{1985}).

\bibitem[{\citenamefont{Walter and Newport}(1997)}]{walter97}
\bibinfo{author}{\bibfnamefont{J.}~\bibnamefont{Walter}} \bibnamefont{and}
  \bibinfo{author}{\bibfnamefont{J.~W.} \bibnamefont{Newport}},
  \bibinfo{journal}{Science} \textbf{\bibinfo{volume}{275}},
  \bibinfo{pages}{993 } (\bibinfo{year}{1997}).

\bibitem[{\citenamefont{Rhind}(2006)}]{rhind06}
\bibinfo{author}{\bibfnamefont{N.}~\bibnamefont{Rhind}}, \bibinfo{journal}{Nat.
  Cell Biol.} \textbf{\bibinfo{volume}{8}}, \bibinfo{pages}{1313}
  (\bibinfo{year}{2006}).

\bibitem[{\citenamefont{Lucas et~al.}(2000)\citenamefont{Lucas,
  Chevrier-Miller, Sogo, and Hyrien}}]{lucas00}
\bibinfo{author}{\bibfnamefont{I.}~\bibnamefont{Lucas}},
  \bibinfo{author}{\bibfnamefont{M.}~\bibnamefont{Chevrier-Miller}},
  \bibinfo{author}{\bibfnamefont{J.~M.} \bibnamefont{Sogo}}, \bibnamefont{and}
  \bibinfo{author}{\bibfnamefont{O.}~\bibnamefont{Hyrien}},
  \bibinfo{journal}{J. Mol. Biol.} \textbf{\bibinfo{volume}{296}},
  \bibinfo{pages}{769} (\bibinfo{year}{2000}).

\bibitem[{\citenamefont{Kolmogorov}(1937)}]{kolmogorov37}
\bibinfo{author}{\bibfnamefont{A.~N.} \bibnamefont{Kolmogorov}},
  \bibinfo{journal}{Bull. Acad. Sc. USSR, Phys. Ser.}
  \textbf{\bibinfo{volume}{1}}, \bibinfo{pages}{335} (\bibinfo{year}{1937}).

\bibitem[{\citenamefont{Johnson and Mehl}(1939)}]{johnson39}
\bibinfo{author}{\bibfnamefont{W.~A.} \bibnamefont{Johnson}} \bibnamefont{and}
  \bibinfo{author}{\bibfnamefont{F.~L.} \bibnamefont{Mehl}},
  \bibinfo{journal}{Trans. AIME} \textbf{\bibinfo{volume}{135}},
  \bibinfo{pages}{416} (\bibinfo{year}{1939}).

\bibitem[{\citenamefont{Avrami}(1939)}]{avrami39}
\bibinfo{author}{\bibfnamefont{M.}~\bibnamefont{Avrami}}, \bibinfo{journal}{J.
  Chem. Phys.} \textbf{\bibinfo{volume}{7}}, \bibinfo{pages}{1103}
  (\bibinfo{year}{1939});
\textbf{\bibinfo{volume}{8}}, \bibinfo{pages}{212}
  (\bibinfo{year}{1940});
\textbf{\bibinfo{volume}{9}}, \bibinfo{pages}{177}
  (\bibinfo{year}{1941}).

\bibitem[{\citenamefont{Sekimoto}(1984)}]{sekimoto84}
\bibinfo{author}{\bibfnamefont{K.}~\bibnamefont{Sekimoto}},
  \bibinfo{journal}{Physica A} \textbf{\bibinfo{volume}{125}},
  \bibinfo{pages}{261} (\bibinfo{year}{1984});
\textbf{\bibinfo{volume}{135}},
  \bibinfo{pages}{2} (\bibinfo{year}{1986}).

\bibitem[{\citenamefont{Sekimoto}(1991)}]{sekimoto91}
\bibinfo{author}{\bibfnamefont{K.}~\bibnamefont{Sekimoto}},
  \bibinfo{journal}{Int. J. Mod. Phys. B} \textbf{\bibinfo{volume}{5}},
  \bibinfo{pages}{1843} (\bibinfo{year}{1991}).

\bibitem[{\citenamefont{Ben-Naim and Krapivsky}(1996)}]{ben-naim96}
\bibinfo{author}{\bibfnamefont{E.}~\bibnamefont{Ben-Naim}} \bibnamefont{and}
  \bibinfo{author}{\bibfnamefont{P.~L.} \bibnamefont{Krapivsky}},
  \bibinfo{journal}{Phys. Rev. E} \textbf{\bibinfo{volume}{54}},
  \bibinfo{pages}{3562 } (\bibinfo{year}{1996}).

\bibitem[{sta()}]{stall_fork}
\bibinfo{note}{It is known that when DNA polymerases introduce copying errors
  in the process of synthesizing nucleotides, replication forks stall to repair
  the mispairing \cite{lodish08}. These fork stalls occur stochastically in
  time. Thus, in our model, a constant fork velocity is a mean-field
  approximation of the time-varying fork velocities and the fork-to-fork
  variation in progression speed.}

\bibitem[{\citenamefont{Marheineke and Hyrien}(2004)}]{marheineke04}
\bibinfo{author}{\bibfnamefont{K.}~\bibnamefont{Marheineke}} \bibnamefont{and}
  \bibinfo{author}{\bibfnamefont{O.}~\bibnamefont{Hyrien}},
  \bibinfo{journal}{J. Biol. Chem.} \textbf{\bibinfo{volume}{279}},
  \bibinfo{pages}{28071} (\bibinfo{year}{2004}). \bibinfo{note}{To test whether
  replacing the observed decreasing fork velocity with its average is a valid
  approximation, we simulated the two cases with a modified version of the
  phantom-nuclei algorithm discussed in Sec.~\ref{sec:model}. We found that the
  tails of the coalescence distributions for the two cases agree to $\approx$
  1\%, indicating that the two cases also have similar end-time distributions.}

\bibitem[{\citenamefont{Conti et~al.}(2007)\citenamefont{Conti, Sacc\`{a},
  Herrick, Lalou, Pommier, and Bensimon}}]{conti07}
\bibinfo{author}{\bibfnamefont{C.}~\bibnamefont{Conti}},
  \bibinfo{author}{\bibfnamefont{B.}~\bibnamefont{Sacc\`{a}}},
  \bibinfo{author}{\bibfnamefont{J.}~\bibnamefont{Herrick}},
  \bibinfo{author}{\bibfnamefont{C.}~\bibnamefont{Lalou}},
  \bibinfo{author}{\bibfnamefont{Y.}~\bibnamefont{Pommier}}, \bibnamefont{and}
  \bibinfo{author}{\bibfnamefont{A.}~\bibnamefont{Bensimon}},
  \bibinfo{journal}{Mol. Bio. Cell} \textbf{\bibinfo{volume}{18}},
  \bibinfo{pages}{3059} (\bibinfo{year}{2007}).

\bibitem[{\citenamefont{Krauth}(2006)}]{krauth06}
\bibinfo{author}{\bibfnamefont{W.}~\bibnamefont{Krauth}},
  \emph{\bibinfo{title}{Statistical Mechanics: Algorithms and Computations}}
  (\bibinfo{publisher}{Oxford University Press}, \bibinfo{address}{Oxford},
  \bibinfo{year}{2006}).

\bibitem[{\citenamefont{Press et~al.}(2007)\citenamefont{Press, Teukolsky,
  William~T, and Flannery}}]{press07}
\bibinfo{author}{\bibfnamefont{W.~H.} \bibnamefont{Press}},
  \bibinfo{author}{\bibfnamefont{S.~A.} \bibnamefont{Teukolsky}},
  \bibinfo{author}{\bibfnamefont{V.}~\bibnamefont{William~T}},
  \bibnamefont{and} \bibinfo{author}{\bibfnamefont{B.~P.}
  \bibnamefont{Flannery}}, \emph{\bibinfo{title}{Numerical Recipes: The Art of
  Scientific Computing}} (\bibinfo{publisher}{Cambridge University Press},
  \bibinfo{address}{New York}, \bibinfo{year}{2007}), \bibinfo{edition}{3rd}
  ed.

\bibitem[{igo(2008)}]{igor}
\bibinfo{journal}{Wavemetrics Inc.}  (\bibinfo{year}{2008}),
  \urlprefix\url{http://www.wavemetrics.com/} .

\bibitem[{\citenamefont{Gumbel}(1958)}]{gumbel58}
\bibinfo{author}{\bibfnamefont{E.~J.} \bibnamefont{Gumbel}},
  \emph{\bibinfo{title}{Statistics of Extremes}} (\bibinfo{publisher}{Columbia
  Univ. Press}, \bibinfo{address}{New York}, \bibinfo{year}{1958}).

\bibitem[{\citenamefont{Kotz and Nadarajah}(2000)}]{kotz00}
\bibinfo{author}{\bibfnamefont{S.}~\bibnamefont{Kotz}} \bibnamefont{and}
  \bibinfo{author}{\bibfnamefont{S.}~\bibnamefont{Nadarajah}},
  \emph{\bibinfo{title}{Extreme Value Distributions}}
  (\bibinfo{publisher}{Imperial College Press}, \bibinfo{address}{London},
  \bibinfo{year}{2000}).

\bibitem[{\citenamefont{Reiss and Thomas}(2001)}]{reiss01}
\bibinfo{author}{\bibfnamefont{R.-D.} \bibnamefont{Reiss}} \bibnamefont{and}
  \bibinfo{author}{\bibfnamefont{M.}~\bibnamefont{Thomas}},
  \emph{\bibinfo{title}{Statistical Analysis of Extreme Values}}
  (\bibinfo{publisher}{Birkhauser Verlag}, \bibinfo{address}{Basel},
  \bibinfo{year}{2001}), \bibinfo{edition}{2nd} ed.

\bibitem[{\citenamefont{Caldarelli et~al.}(1996)\citenamefont{Caldarelli,
  Tolla, and Petri}}]{caldarelli96}
\bibinfo{author}{\bibfnamefont{G.}~\bibnamefont{Caldarelli}},
  \bibinfo{author}{\bibfnamefont{F.~D.} \bibnamefont{Di Tolla}},
  \bibnamefont{and} \bibinfo{author}{\bibfnamefont{A.}~\bibnamefont{Petri}},
  \bibinfo{journal}{Phys. Rev. Lett.} \textbf{\bibinfo{volume}{77}},
  \bibinfo{pages}{2503} (\bibinfo{year}{1996}).

\bibitem[{\citenamefont{Moreira et~al.}(2002)\citenamefont{Moreira, Jos\'{e}
  S.~Andrade, and Amaral}}]{moreira02}
\bibinfo{author}{\bibfnamefont{A.~A.} \bibnamefont{Moreira}},
  \bibinfo{author}{\bibfnamefont{J. S.}~\bibnamefont{Andrade}},
  \bibnamefont{and} \bibinfo{author}{\bibfnamefont{L.~A.}
  \bibnamefont{Nunes Amaral}}, \bibinfo{journal}{Phys. Rev. Lett.}
  \textbf{\bibinfo{volume}{89}}, \bibinfo{pages}{268703}
  (\bibinfo{year}{2002}).

\bibitem[{\citenamefont{Tijms}(2004)}]{tijms04}
\bibinfo{author}{\bibfnamefont{H.}~\bibnamefont{Tijms}},
  \emph{\bibinfo{title}{Understanding Probability: Chance Rules in Everyday
  Life}} (\bibinfo{publisher}{Cambridge University Press},
  \bibinfo{address}{Cambridge}, \bibinfo{year}{2004}).

\bibitem[{notc({\natexlab{a}})}]{note_bilinear}
\bibinfo{note}{The original inference of the \textit{in vitro} data was fitted
  using a bi-linear function \cite{herrick02}. Using the transformation in
  Sec.~\ref{sec:exp} to transpose the bi-linear $I_{vitro}$, we obtained a
  scaled bi-linear $I_{vivo}$. For ease of calculation, in parts of the paper,
  we approximate both bi-linear functions with power-law functions. Both
  initiation functions are approximately quadratic ($I_{vitro}\sim t^{2.62}$
  and $I_{vivo}\sim t^{2.45}$)}.

\bibitem[{notd({\natexlab{b}})}]{note_rate_decay}
\bibinfo{note}{$I_{vitro}$ is approximately quadratic except for a decrease
  [$I(t) \rightarrow 0$] toward the end of S phase \cite{herrick02}. Since the
  errors on the decreasing part of the data are large, we neglected the
  decrease in our previous work \cite{jun05a, bechhoefer07}. However, a recent
  repeat of the original experiment has shown that the decrease is not an
  artifact of poor statistics but represents a real feature of the replication
  process \cite{hyrien08_private}. Thus, we examined how the decrease in $I(t)$
  affects the end-time distribution. We found that with the decrease, the mode
  of the distribution was delayed by $\approx$ 0.3\% and the width increased by
  15\%. These quantities approximately translate into an S phase that is
  prolonged by $\approx$ 0.5 min. This difference is small compared to the
  overall duration of S phase (20 min).}

\bibitem[{\citenamefont{Sethna}(2006)}]{Sethna06}
\bibinfo{author}{\bibfnamefont{J.~P.} \bibnamefont{Sethna}},
  \emph{\bibinfo{title}{Statistical Mechanics: Entropy, Order Parameters, and
  Complexity}} (\bibinfo{publisher}{Oxford University Press},
  \bibinfo{address}{Oxford}, \bibinfo{year}{2006}). \bibinfo{note}{The
  derivation of the Gumbel distribution appears in a web-supplement},
  \urlprefix\url{http://pages.physics.cornell.edu/sethna/StatMech/NewExercises%
.pdf} .

\bibitem[{\citenamefont{Thi\'{e}baud and Fischberg}(1977)}]{thiebaud77}
\bibinfo{author}{\bibfnamefont{C.~H.} \bibnamefont{Thi\'{e}baud}}
  \bibnamefont{and}
  \bibinfo{author}{\bibfnamefont{M.}~\bibnamefont{Fischberg}},
  \bibinfo{journal}{Chromosoma} \textbf{\bibinfo{volume}{59}},
  \bibinfo{pages}{253} (\bibinfo{year}{1977}).

\bibitem[{\citenamefont{Bell and Dutta}(2002)}]{bell02}
\bibinfo{author}{\bibfnamefont{S.~P.} \bibnamefont{Bell}} \bibnamefont{and}
  \bibinfo{author}{\bibfnamefont{A.}~\bibnamefont{Dutta}},
  \bibinfo{journal}{Annu. Rev. Biochem.} \textbf{\bibinfo{volume}{71}},
  \bibinfo{pages}{333} (\bibinfo{year}{2002}).

\bibitem[{\citenamefont{Kimmel and Axelrod}(2002)}]{kimmel02}
\bibinfo{author}{\bibfnamefont{M.}~\bibnamefont{Kimmel}} \bibnamefont{and}
  \bibinfo{author}{\bibfnamefont{D.~E.} \bibnamefont{Axelrod}},
  \emph{\bibinfo{title}{Branching Processes in Biology}}
  (\bibinfo{publisher}{Springer-Verlag}, \bibinfo{address}{New York},
  \bibinfo{year}{2002}).

\bibitem[{not({\natexlab{c}})}]{note_low_F}
\bibinfo{note}{In \cite{bechhoefer07}, we estimated a failure rate $F \approx
  10^{-4}$ using a simple model that neglected the complications due to the
  branching process. Since $t^{**}$ depends on the logarithm of $F$ from
  Eq.~\ref{eq:t**tot*}, the factor of 30 between the two estimates of $F$
  results only in a roughly 1 min shift in the $t^{**}$ of the end-time
  distribution and does not alter the qualitative conclusion of the previous
  work.}

\bibitem[{\citenamefont{Spall}(2003)}]{spall03}
\bibinfo{author}{\bibfnamefont{J.~C.} \bibnamefont{Spall}},
  \emph{\bibinfo{title}{Introduction to Stochastic Search and Optimization}}
  (\bibinfo{publisher}{J. Wiley}, \bibinfo{address}{New Jersey},
  \bibinfo{year}{2003}).

\bibitem[{\citenamefont{Moore et~al.}(1999)\citenamefont{Moore, Yang, Truant,
  and Kornbluth}}]{moore99}
\bibinfo{author}{\bibfnamefont{J.~D.} \bibnamefont{Moore}},
  \bibinfo{author}{\bibfnamefont{J.}~\bibnamefont{Yang}},
  \bibinfo{author}{\bibfnamefont{R.}~\bibnamefont{Truant}}, \bibnamefont{and}
  \bibinfo{author}{\bibfnamefont{S.}~\bibnamefont{Kornbluth}},
  \bibinfo{journal}{J. Cell Biol.} \textbf{\bibinfo{volume}{144}},
  \bibinfo{pages}{213} (\bibinfo{year}{1999}).

\bibitem[{\citenamefont{Nguyen et~al.}(2000)\citenamefont{Nguyen, Co, Irie, and
  Lib}}]{nguyen00}
\bibinfo{author}{\bibfnamefont{V.~Q.} \bibnamefont{Nguyen}},
  \bibinfo{author}{\bibfnamefont{C.}~\bibnamefont{Co}},
  \bibinfo{author}{\bibfnamefont{K.}~\bibnamefont{Irie}}, \bibnamefont{and}
  \bibinfo{author}{\bibfnamefont{J.~J.} \bibnamefont{Lib}},
  \bibinfo{journal}{Current Biol.} \textbf{\bibinfo{volume}{10}},
  \bibinfo{pages}{195} (\bibinfo{year}{2000}).

\bibitem[{\citenamefont{Postow et~al.}(2001)\citenamefont{Postow, Crisona,
  Peter, Hardy, and Cozzarelli}}]{postow01}
\bibinfo{author}{\bibfnamefont{L.}~\bibnamefont{Postow}},
  \bibinfo{author}{\bibfnamefont{N.~J.} \bibnamefont{Crisona}},
  \bibinfo{author}{\bibfnamefont{B.~J.} \bibnamefont{Peter}},
  \bibinfo{author}{\bibfnamefont{C.~D.} \bibnamefont{Hardy}}, \bibnamefont{and}
  \bibinfo{author}{\bibfnamefont{N.~R.} \bibnamefont{Cozzarelli}},
  \bibinfo{journal}{Proc. Natl. Acad. Sci.} \textbf{\bibinfo{volume}{98}},
  \bibinfo{pages}{8219–} (\bibinfo{year}{2001}).

\bibitem[{\citenamefont{Skogestad and Postlethwaite}(2005)}]{skogestad05}
\bibinfo{author}{\bibfnamefont{S.}~\bibnamefont{Skogestad}} \bibnamefont{and}
  \bibinfo{author}{\bibfnamefont{I.}~\bibnamefont{Postlethwaite}},
  \emph{\bibinfo{title}{Multivariable Feedback Control}}
  (\bibinfo{publisher}{J. Wiley}, \bibinfo{address}{Chichester},
  \bibinfo{year}{2005}), \bibinfo{edition}{2nd} ed.

\bibitem[{hyr()}]{hyrien08_private}
\bibinfo{note}{O. Hyrien (private communication).}

\bibitem[{\citenamefont{Meister et~al.}(2007)\citenamefont{Meister, Taddei,
  Ponti, Baldacci, and Gasser}}]{meister07}
\bibinfo{author}{\bibfnamefont{P.}~\bibnamefont{Meister}},
  \bibinfo{author}{\bibfnamefont{A.}~\bibnamefont{Taddei}},
  \bibinfo{author}{\bibfnamefont{A.}~\bibnamefont{Ponti}},
  \bibinfo{author}{\bibfnamefont{G.}~\bibnamefont{Baldacci}}, \bibnamefont{and}
  \bibinfo{author}{\bibfnamefont{S.~M.} \bibnamefont{Gasser}},
  \bibinfo{journal}{EMBO J.} \textbf{\bibinfo{volume}{26}},
  \bibinfo{pages}{1315–} (\bibinfo{year}{2007}).

\bibitem[{\citenamefont{Jun et~al.}(2004)\citenamefont{Jun, Herrick, Bensimon,
  and Bechhoefer}}]{jun04}
\bibinfo{author}{\bibfnamefont{S.}~\bibnamefont{Jun}},
  \bibinfo{author}{\bibfnamefont{J.}~\bibnamefont{Herrick}},
  \bibinfo{author}{\bibfnamefont{A.}~\bibnamefont{Bensimon}}, \bibnamefont{and}
  \bibinfo{author}{\bibfnamefont{J.}~\bibnamefont{Bechhoefer}},
  \bibinfo{journal}{Cell Cycle} \textbf{\bibinfo{volume}{3}},
  \bibinfo{pages}{223} (\bibinfo{year}{2004}).

\bibitem[{\citenamefont{Lodish et~al.}(2008)\citenamefont{Lodish, Berk, Kaiser,
  Krieger, Scott, Bretscher, Ploegh, and Matsudairu}}]{lodish08}
\bibinfo{author}{\bibfnamefont{H.}~\bibnamefont{Lodish}},
  \bibinfo{author}{\bibfnamefont{A.}~\bibnamefont{Berk}},
  \bibinfo{author}{\bibfnamefont{C.~A.} \bibnamefont{Kaiser}},
  \bibinfo{author}{\bibfnamefont{M.}~\bibnamefont{Krieger}},
  \bibinfo{author}{\bibfnamefont{M.~P.} \bibnamefont{Scott}},
  \bibinfo{author}{\bibfnamefont{A.}~\bibnamefont{Bretscher}},
  \bibinfo{author}{\bibfnamefont{H.}~\bibnamefont{Ploegh}}, \bibnamefont{and}
  \bibinfo{author}{\bibfnamefont{P.}~\bibnamefont{Matsudairu}},
  \emph{\bibinfo{title}{Molecular Cell Biology}} (\bibinfo{publisher}{W.H.
  Freeman and Company}, \bibinfo{address}{New York}, \bibinfo{year}{2008}),
  \bibinfo{edition}{6th} ed.

\bibitem[{\citenamefont{Gilbert}(2001)}]{gilbert01}
\bibinfo{author}{\bibfnamefont{D.~M.} \bibnamefont{Gilbert}},
  \bibinfo{journal}{Science} \textbf{\bibinfo{volume}{294}},
  \bibinfo{pages}{96} (\bibinfo{year}{2001}).

\bibitem[{\citenamefont{Anglana et~al.}(2003)\citenamefont{Anglana, Apiou,
  Bensimon, and Debatisse}}]{anglana03}
\bibinfo{author}{\bibfnamefont{M.}~\bibnamefont{Anglana}},
  \bibinfo{author}{\bibfnamefont{F.}~\bibnamefont{Apiou}},
  \bibinfo{author}{\bibfnamefont{A.}~\bibnamefont{Bensimon}}, \bibnamefont{and}
  \bibinfo{author}{\bibfnamefont{M.}~\bibnamefont{Debatisse}},
  \bibinfo{journal}{Cell} \textbf{\bibinfo{volume}{114}}, \bibinfo{pages}{385}
  (\bibinfo{year}{2003}).

\bibitem[{\citenamefont{Patel et~al.}(2006)\citenamefont{Patel, Arcangioli,
  Baker, Bensimon, and Rhind}}]{patel06}
\bibinfo{author}{\bibfnamefont{P.~K.} \bibnamefont{Patel}},
  \bibinfo{author}{\bibfnamefont{B.}~\bibnamefont{Arcangioli}},
  \bibinfo{author}{\bibfnamefont{S.~P.} \bibnamefont{Baker}},
  \bibinfo{author}{\bibfnamefont{A.}~\bibnamefont{Bensimon}}, \bibnamefont{and}
  \bibinfo{author}{\bibfnamefont{N.}~\bibnamefont{Rhind}},
  \bibinfo{journal}{Mol. Biol. Cell} \textbf{\bibinfo{volume}{17}},
  \bibinfo{pages}{308} (\bibinfo{year}{2006}).

\bibitem[{\citenamefont{Heichinger et~al.}(2006)\citenamefont{Heichinger,
  Penkett, B\"{a}hler, and Nurse}}]{heichinger06}
\bibinfo{author}{\bibfnamefont{C.}~\bibnamefont{Heichinger}},
  \bibinfo{author}{\bibfnamefont{C.~J.} \bibnamefont{Penkett}},
  \bibinfo{author}{\bibfnamefont{J.}~\bibnamefont{B\"{a}hler}},
  \bibnamefont{and} \bibinfo{author}{\bibfnamefont{P.}~\bibnamefont{Nurse}},
  \bibinfo{journal}{EMBO J.} \textbf{\bibinfo{volume}{25}},
  \bibinfo{pages}{5171–} (\bibinfo{year}{2006}).

\end{thebibliography}

%

\end{document}